\documentclass[fleqn,usenatbib]{mnras}
\usepackage{hyperref} %MNRAS
\usepackage{graphicx}
\usepackage[T1]{fontenc} %MNRAS
\usepackage{ae,aecompl}  %MNRAS

\usepackage{color,soul}

\newcommand\un[1]{{\,\rm #1}}
\newcommand\rs[1]{_\mathrm{#1}}
\newcommand\U[1]{{\,\rm #1}}
\newcommand\E[1]{\times10^{#1}}

\title[Radio polarization maps of SNRs]{Radio polarization maps of shell-type SNRs\\
II. Sedov models with evolution of turbulent magnetic field}
\author[O. Petruk, R. Bandiera, V. Beshley, S. Orlando, M. Miceli]{O. Petruk$^{1,2}$, R. Bandiera$^3$, V. Beshley$^{2,4}$, S. Orlando$^1$, M. Miceli$^{5,1}$\\
$^{1}$INAF - Osservatorio Astronomico, Piazza del Parlamento, 1, 90134 Palermo, Italy\\
$^{2}$Institute for Applied Problems in Mechanics and Mathematics, Naukova Str., 3-b, 79060 Lviv, Ukraine\\
$^{3}$INAF - Osservatorio Astrofisico di Arcetri, Largo E. Fermi 5, 50125 Firenze, Italy\\
$^{4}$Astronomical Observatory, National University, Kyryla and Methodia Str., 8, 79005 Lviv, Ukraine\\
$^{5}$Dipartimento di Fisica e Chimica, Universit\'a degli Studi di Palermo, Viale delle Scienze, 17, 90128 Palermo, Italy\\
}

\date{Last updated ...; in original form ...}
\pubyear{2016}

\begin{document}

\label{firstpage}
\pagerange{\pageref{firstpage}--\pageref{lastpage}} 
\maketitle

\begin{abstract}
Polarized radio emission has been mapped with great detail in several Galactic supernova remnants (SNRs), but has not yet been exploited to the extent it deserves. We have developed a method to model maps of the Stokes parameters for shell-like SNRs during their Sedov evolution phase. At first, 3-dimensional structure of a SNR has been computed, by modeling the distribution of the magnetohydrodynamic parameters and of the accelerated particles. The generation and dissipation of the turbulent component of magnetic field everywhere in SNR are also considered taking into account its interaction with accelerated particles. Then, in order to model the emission, we have used a generalization of the classical synchrotron theory, valid for the case in which the magnetic field has ordered and disordered components. Finally, 2-dimensional projected maps have been derived, for different orientations of SNR and of interstellar magnetic field with respect to the observer. An important effect to consider is the Faraday rotation of the polarization planes inside the SNR interior. In this paper we present details of the model, and describe general properties of the images. %\textcolor{green}{obtained by our numerical simulations. One effect is that a non-negligible internal Faraday rotation may be responsible of observed polarization maps that are not strictly related to the actual magnetic field structure in SNR. In addition, the turbulent component of magnetic field decreases the fraction of the polarized emission and increases the synchrotron flux, with respect to the case in which the magnetic field is fully ordered.}
\end{abstract}

\begin{keywords}
(ISM:) supernova remnants -- shock waves -- ISM: cosmic rays
-- radiation mechanisms: non-thermal -- acceleration of particles 
\end{keywords}

%%%%%%%%%%%%%%%%%%%%%%%%%%%%%%%%%%%%%%%%%%%%%%%%%%%%%%%%%%%%%%%%%%%%%%%%%
\section{Introduction}

Supernova remnants (SNRs) are very important objects for High-Energy Astrophysics. They provide experimental information, in particular, on the processes related to magnetic turbulence, cosmic rays (CRs), their interactions and influence on the fluid dynamics and the shock properties. Interstellar magnetic field (ISMF, of order of few $\un{\mu G}$) considerably affects the  propagation of all CRs with energies $\leq 3\E{15}\un{eV}$ deviating them from their original directions. Therefore, the only possibility to study these cosmic accelerators individually is to consider different kinds of emission resulted from interactions of accelerated particles with magnetic fields, with photons, or with other particles. 

SNRs are observed in all electromagnetic domains, from radio waves to TeV gamma-rays. There is a wealth of information produced by many experiments but only a fraction of it has been exploited. 

Fluxes and spectra, both integrated and spatially resolved, are widely used. In some cases, fine details of the SNR surface brightness distribution have been deeply investigated. For instance, the radial thickness of the X-ray SNR rim is associated to the strength of the magnetic field (MF) behind the shock front \citep{vink-laming-2003} if the dominant factor of the fading process is the density decrease and the energy downgrading (and/or diffusion) of the emitting particles  \citep{Ber-Volk-2003-mf,Ber-Volk-2004-mf,Bamba-etal-2005,Volk-Ber-2005-mf,Warren-et-al-2005}. Alternatively, it may indicate the length-scale of the MF damping downstream \citep{Pohl-et-al-2005,Pohl-et-al-2015,Ressler-et-al-2014,Tran-et-al-2015}. 
Additional constraints on models come from the energy dependence of the rim thickness \citep{Ressler-et-al-2014}.
The presence (or absence) of the X-ray emission in the precursor gives hints on the particle acceleration properties \citep{Long-et-al-2003,Morlino-et-al-2010,Winkler-etal-2014}. The azimuthal variation of the radio brightness depends on the obliquity dependence of the injection efficiency \citep{Fulbright-Reynolds-1990,Petruk-et-al-2009} while the X-ray azimuthal profile might help in determining whether the maximum particle momentum is time- or loss-limited \citep{Petruk-et-al-2011}. High spatial resolution observations may reveal the distance between the shock and the contact discontinuity \citep[e.g.][]{Miceli-et-al-2009}. A considerably reduced distance can be effect of particle acceleration with an efficient back-reaction on the flow \citep{Warren-et-al-2005} or otherwise of protrusions of ejecta beyond the forward shock \citep{Rakowski-et-al-2011}. Similar effects may be due to the ejecta clumping \citep{Orlando-etal-2012}. 
Rapidly varying spots in the SNR shell might be a sign of the presence of either a highly amplified MF and ongoing acceleration \citep[e.g.][]{Uchiyama-et-al-2007} or of the plasma waves and steady electron distributions \citep{Bykov-et-al-2008,Bykov-et-al-2009}. The ordered stripes observed in Tycho SNR not far from the shell \citep{Eriksen-et-al-2011} are a challenge for models of plasma microphysics \citep{Bykov-et-al-2011,Malkov-et-al-2012,Laming-2015}. 

In contrast to the widespread analysis of spectra and local spatial characteristics of the nonthermal emission from SNRs in different spectral ranges, the maps of the surface brightness and of the polarized synchrotron emission -- which are known for many SNRs -- are lacking an adequate analysis. Such situation should be changed by developing new analysis methods. In general, there are two ways to deal with SNR images: to analyze observed maps with minimal assumptions or to model maps starting from basic theoretical principles. 

Using the properties of the emission processes and the observed maps in different bands it is possible, for example, to separate the thermal and nonthermal X-ray images out of the mixed observed one \citep{Miceli-et-al-2009}, to predict a gamma-ray image of SNR \citep{Petruk-etal-2009c} or to determine the MF strength in the limbs of SNR \citep{Petruk-etal-2012}. 

As to the latter strategy to explore SNR images, a method to model the synchrotron radio and X-ray images of Sedov SNRs (that is the shell-like adiabatic remnant of a spherical explosion in a uniform medium) was developed by \citet{Fulbright-Reynolds-1990} and \citet{Reyn-98}, who also analyzed the main features of the these maps. The method was adapted to the gamma-ray SNR images due to the inverse-Compton emission \citep{Petruk-et-al-2009b,xmaps} and to those due to the hadronic interactions with protons inside SNR \citep{ppmaps}. The method was generalized to the case of a SNR evolving in an ISM with nonuniform distributions of density and/or MF: the resulting asymmetries on the radio maps have been studied by \citet{Orlando-etal-2007} and in X-rays and $\gamma$-rays by \citet{Orlando-etal-2011}. 

The SNRs morphologies shown by their maps have the potential for understanding the plasma microphysics \citep{Reynolds-2004} and determining the properties of the SNR and its environment. 
In particular, they allow for determination of the 3-dimensional orientation of ISMF and even its gradient around SN~1006 \citep{Petruk-et-al-2009,Bocchino-et-al-2011} or, by analyzing a sample of SNRs, to reveal properties of the Galactic MF \citep{West-SafiHarb-et-al-2016}. 
By modeling the surface brightness maps, it was shown that a certain class of SNRs, namely the thermal X-ray composites (or mixed-morphology SNRs, i.e. remnants with a thermal X-ray peak within the radio shell), might be explained as a projection effect of SNR evolving in essentially nonuniform environment \citep{petruk-2001}. 

The work of \citet{band-petr-2016a} (hereafter Paper I) and the present paper continue a series of articles devoted to development of methods to model SNR images. Namely, we are interested in maps of the surface distribution of the polarized radio emission that could also be important to test theories. 
It is known that random MF generally reduces the polarization of the synchrotron emitting sources \citep[e.g.][]{Stroman-Pohl-2009}, but it is able also to produce stochastic small-scale patchy structures, like spots and filaments, with high polarization fraction in synchrotron X-rays \citep{Bykov-et-al-2008,Bykov-et-al-2009}.
Recently, \citet{Schneiter-etal-2015} have considered a task to simulate polarization images of SNR as a whole under two simplifications: the authors neglected the turbulent component of MF and assumed the (variable) Faraday effect in the SNR interior to be minor comparing to the (uniform) Faraday rotation of the polarization planes outside. In our model, we consider both these components and demonstrate that they might essentially modify polarization images comparing to cases with vanishing turbulent MF and internal Faraday rotation. 

With Paper I, we have essentially reached two main achievements. First, we have developed a method to evaluate the synchrotron polarized emission in the case in which the MF is the composition of an uniform plus a random component, and we have computed exact analytic formulae for all Stokes parameters in some particular cases, the most useful probably being that of a power-law particle energy distribution and a Gaussian isotropic MF random component: the resulting formulae show strong similarities with those for the classical case (i.e. of a homogeneous MF), and are in fact a generalization of them. Then, using a thin-shell approximation, we have simulated the observed maps of the total radio intensity and of the Stokes parameters for a wide number of cases. In spite of the simplification used, we have been able to tackle in detail a number of effects, like for instance the effectiveness of the internal Faraday rotation in distorting the polarization pattern as well as on affecting the observed polarization fraction; and the partial depolarization also due to projection effects and to the random MF component. Another advantage of that approach is that it is computationally rather light.

In the present paper we present instead a more detailed model; namely, we consider the distribution of polarized synchrotron emission not only in the thin shell but everywhere inside a shell-like SNR. In this way more reliable maps can be obtained, which may be used in particular to test various effects as well as to verify how accurate the simplified models of Paper I can be, under typical conditions. 

%%%%%%%%%%%%%%%%%%%%%%%%%%%%%%%%%%%%%%%%%%%%%%%%%%%%%%%%%%%%%%%%%%%%%%%%%
\section{Polarization in adiabatic SNR}

In the present paper, aimed at modeling the radio emission from shell-type SNRs, we have improved the treatment presented in Paper I, by implementing a detailed and self-consistent treatment of SNRs in the adiabatic regime, valid for objects old enough to have swept up from the ambient medium a mass much larger than the original mass of the supernova ejecta, but that are still evolving adiabatically, i.e. the radiative losses and the back reaction of accelerated particles are inefficient \citep[in fact, the SNR statistics is consistent with their radio emission to be strongly reduced about the end of the adiabatic phase;][]{band-petr-2010}. 
The approach we present here  (Sects.~\ref{polariz:sect2.2}, \ref{polariz:sect2.3}, \ref{polariz:sect2.4}, \ref{polariz:eqwaves}) is general, in the sense that it may be used 
with the numerical 3D MHD simulations of an adiabatic remnant of an asymmetric supernova explosion in a medium with nonuniform distributions of density and/or magnetic field. However, in order to simplify the analysis, to make conclusions clearer and to have a basic reference model, in the present paper, we apply our model to the so-called Sedov SNR, i.e. the remnant of a spherical point-like explosion in the uniform medium (Sect.~\ref{polariz:sect2.1}).  Polarization maps of SNRs from fully numerical 3-D MHD simulations with asymmetric ejecta and/or expansion  in a nonuniform environment are in preparation.

\subsection{Magnetohydrodynamics and relativistic electrons in Sedov SNRs}
\label{polariz:sect2.1}

{Before calculations of the polarization maps, one has to simulate the 3D MHD structure of an SNR.}
 
{In the simulations reported in} the present paper, we assume a strong unmodified shock (with $\gamma=5/3$), {adiabatically} expanding in a uniform medium, after a spherical point-like explosion. This problem admits an analytic solution \citep{Sedov-59}, which is self-similar: namely, the spatial and temporal variation of parameters inside the SNR, for example the flow velocity $u(r,t)$, can be written in the form
\begin{equation}
 u(r,t)=u\rs{s}(t)\,\bar u(\bar r),\quad \bar r=r/R(t)
\end{equation}
where the index `s' corresponds to the immediately post-shock position and the profile $\bar u(\bar r)$ does not explicitly depend on time. In other words, it is sufficient to calculate once the normalized profile $\bar u(\bar r)$, to obtain (just by scaling) the profile at any given time. The scaling properties of the Sedov solution reduce the number of free parameters in simulations and reveal how the supernova and the environment characteristics combine.\footnote{{The Sedov solution describes the remnant of a point-like explosion and therefore neglects the structure of ejecta which may give some effect in the polarization maps at the early stages of SNR evolution.}}
Accurate approximations of the Sedov solution have been proposed for calculations, either in the Eulerian coordinate $r$ \citep{Cox-Franko-1981} or in the Lagrangian coordinate $a$ \citep[Sect.~4.5 in][]{approxSedov}.\footnote{The Lagrangian approach is useful if one needs to trace the evolution of parameters inside a given fluid parcel. $r$ is the common spatial coordinate, while $a$ is like a number attached to a fluid element. It is defined as the radial coordinate of the element before the shock crosses this element and remains unchanged later; i.e. $a\equiv R(t\rs{i})$ where $t\rs{i}$ is the time  when the fluid element $a$ was shocked. There is a possibility to convert $a$ to $r$ if one knows the hydrodynamical structure downstream. Such conversion is a way to find a spatial distribution of parameters.} 

The model assumes that the MF is a composition of an ordered ($B$) and a disordered ($\delta B$) component. The distribution of $B$ inside the SNR may be calculated following \citet{Chevalier1974}: the MF in each point may be treated as the sum of a radial and a tangential component, which evolve in different ways \citep[for detailed references and expressions see Sect.~2.2 in][]{petr-2016}. In a Sedov SNR, like for the other hydrodynamic parameters, also the spatial profiles of these two quantities might be expressed with self-similar functions of the normalized coordinates $\bar r$ or $\bar a=a/R$. The MF experiences a compression at the shock, with the compression factor $\sigma\rs{B}={B\rs{s}}/{B\rs{o}}$ which depends on the shock obliquity angle $\Theta\rs{o}$ (the angle between the ambient MF and the shock normal) according to the law:
\begin{equation}
 \sigma\rs{B}=\left(\cos^2\Theta\rs{o}+\sigma^2 \sin^2\Theta\rs{o}\right)^{1/2},
\end{equation}
where the index `o' refers to the pre-shock position, and $\sigma=\rho\rs{s}/\rho\rs{o}$ is the (density) shock compression factor.
Consistently with our assumptions, we will use $\sigma=4$ in our calculations. Therefore, $\sigma\rs{B}$ runs from $1$, for a parallel shock ($\Theta\rs{o}=0^\mathrm{o}$), to $4$, for a perpendicular shock ($\Theta\rs{o}=90^\mathrm{o}$).
Instead, the model for evolution of the disordered MF component will be presented in Sect.~\ref{polariz:eqwaves}. 

In order to synthesize the SNR synchrotron maps one has to know also the spectral and spatial distributions of relativistic electrons. We assume the power-law spectrum of emitting particles with the index $s=2$ and the injection efficiency (defined as the fraction of accelerated particles) independent of the shock obliquity (unless otherwise stated). The evolution of the relativistic electron population downstream of the shock is also self-similar; we model it following \citet{Reyn-98}; details are presented in \citet[][Appendix A]{xmaps}.

\subsection{Polarized emission}
\label{polariz:sect2.2}

Once the distributions of MF and emitting particles have been evaluated for all locations within the SNR, we can compute the local Stokes parameters, ${\cal I}$, ${\cal Q}$, and ${\cal U}$, which contain full information on the (linearly) polarized synchrotron emission per unit path; while ${\cal V}$, the parameter associated to circular polarization, vanishes.
We can then integrate them along the line of sight, to get the observed Stokes parameters.

For a general orientation $x$-$y$ of the projected axes {($z$ is the line-of-sight coordinate with the positive direction toward the observer)}, the standard formulae give ${\cal Q}=\langle E\rs{x}E^\mathrm{*}\rs{x}\rangle-\langle E\rs{y}E^\mathrm{*}\rs{y}\rangle$ and  ${\cal U}=\langle E\rs{x}E^\mathrm{*}\rs{y}\rangle+\langle E\rs{y}E^\mathrm{*}\rs{x}\rangle$.
If we choose the special orientation ($x'$-$y'$), such that the $E'$ vector is directed towards the unit vector $\hat x'$ (namely the $B'$ vector is directed towards the unit vector $\hat y'$), we readily derive ${\cal Q}'\geq0$ and ${\cal U}'=0$.

\begin{figure}
 \centering
 \includegraphics[width=8.4truecm]{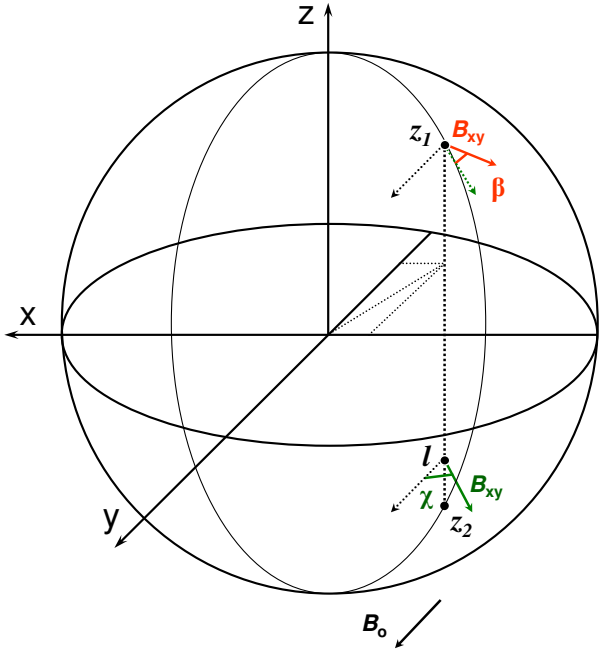}
 %\vspace{5truecm}
 \caption{MF orientation as it would be derived from the polarized emission. {OZ axis points toward the observer.} Line of sight is parallel to OZ axis. Emission risen at point $l$, with orientation of the ordered component of MF shown by the green vector (lower right), changes its polarization plane inside SNR, on the way toward the observer, due to the Faraday rotation inside SNR. It leaves SNR at point $z_1$. `MF orientation' inferred from this ``portion of emission'' is shown by the red vector (upper right).  The angle $\chi$ determines the orientation of the ordered component of MF in each point while the rotation on the angel $\beta$ is due to the Faraday effect.
               }
 \label{polariz:fig_pic}
\end{figure}

Let $P_{\perp}$ and $P_{\|}$ be, respectively, the synchrotron power per unit frequency $\nu$ polarized perpendicular and parallel to the projected MF direction, then ${\cal I}'=(P_{\perp}+P_{\|})/4\pi$ and ${\cal Q}'=(P_{\perp}-P_{\|})/4\pi$.
For an arbitrary orientation, the Stokes parameters can be derived using the standard rotation laws:
\begin{equation}
 \left\{
 \begin{array}{lcl}
  {\cal I}&=&{\cal I}'\\
  {\cal Q}&=&{\cal Q}'\cos2\chi-{\cal U}'\sin2\chi={\cal Q}'\cos2\chi\\
  {\cal U}&=&{\cal Q}'\sin2\chi+{\cal U}'\cos2\chi={\cal Q}'\sin2\chi\\
  {\cal V}&=&{\cal V}'=0
 \end{array}
 \right.
\end{equation}
where $\chi$ is an angle between the $x$ axis and the local polarization measured counter-clockwise, from the observer point of view. This is the same as an angle  
between the $y$ and the local vector $\mathbf{B}\rs{xy}$ (a component of the ordered MF $\mathbf{B}$ in the plane orthogonal to the line of sight\footnote{In the present paper, $\chi$ is the rotation angle of the MF vector while in the Paper I it was the rotation angle of the coordinate system. 
The sign in this definition is opposite to that in Paper I, and now gives an orientation consistent with the angle $\beta$, for the Faraday rotation. It also applies to (\ref{polariz:Stokes-par-project}), cf. Eqs.~(10) and (69) in Paper I (see Fig.~\ref{polariz:fig_pic}). On the other hand, this change of definition does not affect the results.}).

In terms of the components $B\rs{x}$ and $B\rs{y}$ of the ordered projected MF, we get $\cos\chi=B\rs{y}/B\rs{xy}$, $\sin\chi=-B\rs{x}/B\rs{xy}$,
where $B\rs{xy}^2=B\rs{x}^2+B\rs{y}^2$ (see Appendix~\ref{polariz:appA}). Therefore, 
\begin{equation}
 \left\{
 \begin{array}{lcl}
  \cos2\chi&=&\displaystyle\frac{B\rs{y}^2-B\rs{x}^2}{B\rs{x}^2+B\rs{y}^2},\\ \\
  \sin2\chi&=&-\displaystyle\frac{2B\rs{x}B\rs{y}}{B\rs{x}^2+B\rs{y}^2}.
 \end{array}
 \right.
 \label{polariz:cossin}
\end{equation}

%--------------------------------------------------------------
%--------------------------------------------------------------

Before integrating the Stokes local values along the line of sight, to obtain observed Stokes parameters, we must account also for the rotation of the polarization plane inside the SNR due to Faraday effect (called `internal Faraday rotation' throughout this paper):
\begin{equation}
 \beta(l,z_1)=RM(l,z_1)\ \lambda^2
 \label{polariz:betadef}
\end{equation}
where $\beta$ is the angle of rotation of the polarization plane during propagation from the emission site $l$ inside the SNR to $z_1$, namely where the line of sight crosses the edge of the SNR near to the observer (see Fig.~\ref{polariz:fig_pic}), measured in the same direction as $\chi$ (i.e. counter-clockwise from the observer point of view); $\lambda$ is the wavelength of emission, and $RM(l,z_1)$ is the rotation measure, defined as:
\begin{equation}
 RM(l,z_1)=\frac{e^3}{2\pi m\rs{e}^2c^4}\int_{l}^{z_1}n\rs{e}B\rs{z}dz.
\label{polariz:RMdef}
\end{equation}
The foreground Faraday rotation (i.e. from the point $z_1$ to the observer location) is not considered in our simulations because our goal is to 
investigate the propagation effects internal to the source (SNR) rather than those in the interstellar medium.

We detect the composition of the emission from each point $l$ along the line of sight, propagated until the point $z_1$.
Therefore, the projected Stokes parameters should be calculated as 
\begin{equation}
 \left\{
 \begin{array}{lcl}
  I&=&\int_{z_2}^{z_1} {\cal I}'(l)\,dl\\ \\
  Q&=&\int_{z_2}^{z_1} {\cal Q}'(l)\cos2\left(\chi(l)+\beta(l,z_1)\right)dl\\ \\
  U&=&\int_{z_2}^{z_1} {\cal Q}'(l)\sin2\left(\chi(l)+\beta(l,z_1)\right)dl
 \end{array}
 \right.
 \label{polariz:Stokes-par-project}
\end{equation}
where $z_2$ is the farther edge of the SNR (Fig.~\ref{polariz:fig_pic}). 

The angle-sum trigonometric identities allow us to use Eqs.~(\ref{polariz:cossin}) and (\ref{polariz:betadef}) here. 

Once we have the projected Stokes parameters calculated for each point of the SNR projection, we may obtain the map of polarization fraction
\begin{equation}
 \Pi=\frac{\sqrt{Q^2+U^2}}{I}
\end{equation}
and of the angle of the `observed' rotation of the polarization plane 
\begin{equation}
 \Psi=\frac{1}{2}\arctan\left(\frac{U}{Q}\right).
\end{equation}
The angle $\Psi$ gives us the orientation of the observed `projected' MF ${\cal B}$ relative to the projection of the ambient MF $\mathbf{B}\rs{o}$:
\begin{eqnarray}
 {\cal B}\rs{x}&\propto& \sin\Psi,\\ 
 {\cal B}\rs{y}&\propto& \cos\Psi. 
\end{eqnarray}

In the discussion of results (Sect.~\ref{polariz:sect-results}), we consider also an average polarization fraction for a SNR image $\langle\Pi\rangle$ which is an arithmetic average of $\Pi$ from each (out of $N$) point over the SNR projection:\footnote{{We would like to make clear that this average polarization fraction is not how observers typically calculate the polarization fraction for an entire object. The standard practice is the total polarized power divided by the total flux density (our Eq.~\ref{polariz:avePi} does not weight regions by total flux). Instead, Eq.~(\ref{polariz:avePi}) gives an average of the local fractions over the image and it is used solely to characterize the dominant values of $\Pi$ in the image, not the object as a whole.}}
\begin{equation} 
 \langle\Pi\rangle=\frac{1}{N}\sum_i^N \Pi_i .
\label{polariz:avePi}
\end{equation}

{It should also be noted that when gradients in Stokes parameters occur, unless they are fully resolved, they will entail an effective `beam depolarization'. That is, the polarized fractions will be resolution-dependent and can be expected to rise with increasing $N$. In our simulations, we use a high resolution, therefore our local polarization fractions may reach the maximum theoretical value. It is important to account for this effect when comparing the simulations with observations.}

\subsection{Stokes parameters. Completely ordered MF}
\label{polariz:sect2.3}

In the classic synchrotron emission theory, the Stokes parameter ${\cal I}'$ for particles with the single momentum $p$ is
\begin{equation}
 {\cal I}'\rs{single}(\nu,p)=\frac{\sqrt{3}\,e^3}{4\pi mc^2}B\rs{xy}F\left(\frac{\nu}{\nu\rs{c}}\right),
\end{equation}
and the parameter ${\cal Q}'$ is 
\begin{equation}
 {\cal Q}'\rs{single}(\nu,p)=\frac{\sqrt{3}\,e^3}{4\pi mc^2}B\rs{xy}G\left(\frac{\nu}{\nu\rs{c}}\right)
\end{equation}
where $B\rs{xy}$ is the component of MF in the projection plane (i.e. perpendicular to the line of sight), $F$ and $G$ are known functions, $\nu\rs{c}$ the critical frequency. 

The Stokes parameters for electrons distributed with the power law in momentum $N(p)=Kp^{-s}$ are given by convolution over $p$: 
\begin{eqnarray}
{\cal I}'&=&\frac{s+7/3}{s+1}\,{\cal C} K B\rs{xy}^{(s+1)/2},	\label{eq:Iordered}\\
{\cal Q}'&=&{\cal C} K B\rs{xy}^{(s+1)/2}, \label{eq:Qordered}
\end{eqnarray}
where
\begin{equation}
 {\cal C}=
 \frac{\sqrt{3}\,e^3}{4\pi mc^2}\left(\frac{\nu}{c_1}\right)^{-(s-1)/2}
 \Gamma\left(\frac{s}{4}+\frac{7}{12}\right)\Gamma\left(\frac{s}{4}-\frac{1}{12}\right),
\end{equation}
and $c_1=3e/(2\pi m^3c^5)$.
In the case of the ordered MF, the polarization fraction is the maximum possible one: 
\begin{equation}
 \Pi\rs{max}=\frac{s+1}{s+7/3}.
\label{polariz:Pimaxdef}
\end{equation}

Results of simulations of the polarized emission from the Sedov SNR with the only ordered MF are presented in Sect.~\ref{polariz:ordfieldresults}.

\subsection{Stokes parameters. Ordered plus disordered MF}
\label{polariz:sect2.4}

If a model considers the only ordered MF then it results in the high fraction of the polarized emission. Observations reveal rather small $\Pi$ in SNRs, around $15\%$. Therefore, a model of polarized synchrotron emission should also  include the disordered component of MF inside SNR. 

At this point we face the two problems. 

%--------------------------------------------------------------
\begin{figure}
 \centering
 \includegraphics[width=8.2truecm]{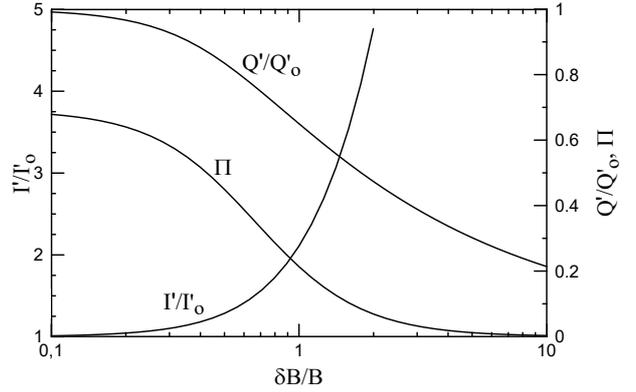}
 %\vspace{5truecm}
 \caption{Dependence of the Stokes parameters ${\cal I}'/{\cal I}'\rs{o}$ and ${\cal Q}'/{\cal Q}'\rs{o}$ as well as the polarization fraction $\Pi$ on the ratio of strengths of the disordered and ordered components of MF $\delta B/B$, for the power-law index of the electron spectrum $s=2$ ($\Pi$ is almost insensitive to $s$; for the ${\cal I}'(s)$, ${\cal Q}'(s)$ dependence see Fig.~1 in Paper I).
               }
 \label{polariz:fig_pol}
\end{figure}
%--------------------------------------------------------------

First, the classic theory of synchrotron emission is developed for MF which is uniform on the lengthscale much greater than the gyroradius of the emitting electron, in other words, it may be used with the ordered MF only. Thus, the theory of synchrotron emission should be generalized to cases when the random MF $\delta B$ is not negligible comparing to the ordered MF $B$. We have developed such generalization in Paper I. Namely, if the particles with the power-law momentum distribution emit in MF which has ordered component of the strength $B$ and the random component represented by the spherical Gaussian\footnote{{
A possible realization of a turbulent MF by a spherical Gaussian, which preserves the zero divergence of MF, by an ensembles of Alfv\'en waves, is given by \citet{Bykov-et-al-2008}.}} 
with the standard deviation $\delta B$ then the Stokes parameters are {(for details see Sect.2.2 in the Paper I)}
\begin{eqnarray}
{\cal I}'\!\!\!\!\!&=&\!\!\!\!\!{\cal I}'\rs{o}\left\{2^{(s+1)/4}\Gamma\left(\frac{s+5}{4}\right)
\right.\nonumber	\\
\label{eq:IPL}
\!\!\!\!\!&\cdot&\!\!\!\!\!\left.
\left(\frac{\delta B}{B}\right)^{(s+1)/2}{_{1}F_{1}}\!\left(-\frac{1+s}{4},1,-\frac{1}{2}\left(\frac{\delta B}{B}\right)^{-2}\right)\right\}
;	\\
{\cal Q}'\!\!\!\!\!&=&\!\!\!\!\!{\cal Q}'\rs{o}\left\{2^{(s-7)/4}\Gamma\left(\frac{s+9}{4}\right)
\right.\nonumber	\\
\label{eq:QPL}
\!\!\!\!\!&\cdot&\!\!\!\!\!\left.
\left(\frac{\delta B}{B}\right)^{(s-3)/2}{_{1}F_{1}}\!\left(\frac{3-s}{4},3,-\frac{1}{2}\left(\frac{\delta B}{B}\right)^{-2}\right)\right\},
\end{eqnarray}
where ${_{1}F_{1}}(a,b,z)$ is the Kummer confluent hypergeometric function, ${\cal I}'\rs{o}$ and ${\cal Q}'\rs{o}$ are respective Stokes parameters for the case of the ordered MF (i.e. given by Eqs.~\ref{eq:Iordered} and \ref{eq:Qordered}).
The polarization fraction is 
\begin{eqnarray}
\Pi\!\!\!\!\!&=&\!\!\!\!\!\Pi\rs{max}\left\{\frac{s+5}{16}\left(\frac{\delta B}{B}\right)^{-2}
\right.\nonumber	\\
&&\qquad \cdot\left.
\frac{{_{1}F_{1}}\!\left((3-s)/4,3,-(\delta B/B)^{-2}/2\right)}{{_{1}F_{1}}\!\left(-(1+s)/4,1,-(\delta B/B)^{-2}/2\right)}\right\}.
\end{eqnarray}
The terms in braces $\{...\}\rightarrow 1$ for $\delta B/B\rightarrow 0$ and the results of the classic synchrotron emission theory are recovered for a vanishing random component of MF. Fig.~\ref{polariz:fig_pol} demonstrates the dependence of ${\cal I}'$, ${\cal Q}'$ and $\Pi$ on the strength of the disordered MF. Note, that ${\cal I}'$ increases with $\delta B/B$; this is of importance for fitting of the observed synchrotron spectra. In particular, if one assumes $\delta B/B\sim 1$ than the flux is twice the flux given by the classic synchrotron theory. {Similar effect, namely, that the turbulent MF is able to enhance the synchrotron emission, is noticed also by \citet{Bykov-et-al-2008}.}

The second problem is that we have to know the value of $\delta B$ everywhere inside SNR. An approach to model the evolution of the disordered component of MF on the shock and in the SNR interior is one of the main goals of the present paper.
It is developed in Sect.~\ref{polariz:eqwaves} where the equation for the evolution of waves is considered. Solutions of this equation are analyzed in Sect.~\ref{polariz:sol-main-eq}. In our approach, the wave properties are related to properties of CRs accelerated on the forward shock and evolved downstream. 

Once $B$ and $\delta B$ are known in each point of SNR, the projected Stokes parameters may be calculated by Eq.~(\ref{polariz:Stokes-par-project}) with ${\cal I}'$ and ${\cal Q}'$ given by Eqs.~(\ref{eq:IPL}) and (\ref{eq:QPL}). Note, that in our approach to the synchrotron emission theory (where the random MF is represented by the spherical Gaussian) neither $\chi$ (spatial structure of MF) nor $\beta$ (internal Faraday rotation) are affected by the disordered MF. 

Results of simulations of the polarized emission from the Sedov SNR including disordered MF are presented in Sect.~\ref{polariz:disordfieldresults}.

%%%%%%%%%%%%%%%%%%%%%%%%%%%%%%%%%%%%%%%%%%%%%%%%%%%%%%%%%%%%%%%%%%%%%%%%%
\section{Modelling disordered component of MF in SNR}
\label{polariz:eqwaves}

One of the basic ingredients needed to calculate detailed synchrotron polarization maps of SNRs is the spatial distribution of the MF disordered component, everywhere inside the SNR. In this section we present a description of the evolution of this MF component: this treatment is rather general, in the sense that it may be applied to a shock moving in a medium with nonuniform distribution of density and/or MF; however, for the calculation of the numerical models presented in this paper, we will apply it only to cases of a SNR Sedov expansion.
The present treatment is based on the following main assumptions:
i. the disordered MF is essentially the effect of the composition of Alfv\'en waves and we are limited therefore to small values of the ratio $\delta B/B$;
ii. these waves, as well as the CRs, behave like ``test-particles'', i.e. they are developed atop of a MHD structure, which is unaffected by them;
iii. MHD instabilities are not taken into consideration once we consider the Sedov SNR;
iv. waves and CRs interact with each other through resonant interactions only;
v. the wave damping is described by a non-linear Landau approach.

More sophisticated descriptions of the random MF component  are present in the literature \citep[e.g.][and others]{Bell-2004,Balsara-Kim-2005,Giacalone-Jokipii-2007,Mizuno-et-al-2011,Guo-et-al-2012,Fraschetti-2013} and should be considered in the future studies. In the present paper, we adopt instead a rather standard and `simple' prescription for the wave evolution. This allows us to reveal the basic trends caused by the disordered MF, as well as the complexity which arises when taking into consideration the random MF. Also the choice of modelling a Sedov SNR in the present paper is in order to provide some basic `reference' results, for a typical evolutionary phase.

We leave to future simulations the task of revealing effects of other, more complex, elements (like MHD instabilities, nonuniform ISM or ISMF, non-adiabatic or CR-modified shocks, other types of waves) in the modelling of polarized synchrotron emission from SNRs.

\subsection{Equation for the downstream evolution of $\delta B$}

The equation that describes the evolution of waves downstream of the shock is \citep{McKenzie-Volk-1982} %caprioli2009
\begin{equation}
 \frac{\partial E\rs{w}}{\partial t}+\frac{\partial F\rs{w}}{\partial r}=
 u\frac{\partial P\rs{w}}{\partial r}+\sigma\rs{w}P\rs{w}-\Gamma\rs{w}P\rs{w}
\label{polariz:eq1}
\end{equation}
where $u$ is the flow velocity in respect to the shock reference frame, 
$E\rs{w}=\delta B^2/4\pi$ is the wave (magnetic and kinetic) energy density, $P\rs{w}=\delta B^2/8\pi$ is the pressure of waves, $\sigma\rs{w}P\rs{w}$ and $\Gamma\rs{w}P\rs{w}$ are the growth and damping rates for waves respectively. Note that the quantities in this equation are not functions of the wave number, but are the integrated ones. 
Let us assume that the energy flux $F\rs{w}$ is dominated by Alfv\'en waves, so that 
\begin{equation}
 F\rs{w}=\frac{\delta B^2}{4\pi}u + \frac{\delta B^2}{8\pi}u.
\end{equation}
It accounts for the purely magnetic (the first term to the right) and kinetic (the second term) contributions \citep{McKenzie-Volk-1982}, 
$v\rs{A}$ is the Alfv\'en velocity 
\begin{equation}
 v\rs{A}=\frac{B}{\sqrt{4\pi \rho}}.
\label{polariz:eq4}
\end{equation}
Therefore, behind the shock, $E\rs{w}=2P\rs{w}$, $F\rs{w}=3uP\rs{w}$ and equation for $P\rs{w}$ becomes
\begin{equation}
 \frac{\partial P\rs{w}}{\partial t}+u\frac{\partial P\rs{w}}{\partial r}+P\rs{w}\frac{3}{2}\frac{\partial u}{\partial r}=
 \frac{1}{2}\left(\sigma\rs{w}P\rs{w}-\Gamma\rs{w}P\rs{w}\right).
\label{polariz:eq1b}
\end{equation}

Alfv\'en waves interact resonantly with Larmor radius of accelerated protons: $k\propto 1/p$ where $k$ is the wavenumber. 
The growth rate is derived by \citet{Skilling1975iii}:
\begin{equation}
 \sigma\rs{w}(k)=\frac{4\pi}{3}\frac{v\rs{A}\left|\hat b\cdot \hat n\right|}{kW(k)}p^4 v \frac{\partial f}{\partial r}
 \label{polariz2:growthSkill}
\end{equation}
where $W(k)$ is the spectrum of the wave energy density, $\hat b$ is the unit vector in the direction of magnetic field \citep{Skilling1975i} and $\hat n$ is the unit vector outward along the radius. These unit vectors are defined in such way in order their dot product to be $\hat b\cdot \hat n = \cos\Theta$ where $\Theta$ is a `local` obliquity angle (between magnetic field and the radial direction). The absolute value of the dot product is taken because the growth rate is positive while the dot product may be negative depending on the local orientations of these vectors. 
The growth term in Eq.~(\ref{polariz:eq1}) is 
\begin{equation}
 \sigma\rs{w}P\rs{w}=\int \sigma\rs{w}(k) W(k)dk
\end{equation}
We therefore integrate (\ref{polariz2:growthSkill}) using the the property $dk/k=-dp/p$ and the definition of the CR pressure 
\begin{equation}
 P\rs{c}=\frac{4\pi}{3} \int p^4 v f \frac{dp}{p},
\end{equation}
and derive that 
\begin{equation}
 \sigma\rs{w} P\rs{w}=v\rs{A}\cos\Theta\frac{\partial P\rs{c}}{\partial r}.
\label{polariz:eq3}
\end{equation}

The damping term consists of the linear and non-linear contributions \citep[e.g.][]{Ptuskin-Zirak-2003}: $\Gamma\rs{w}=\Gamma\rs{L}+\Gamma\rs{NL}$. The first one is due to collisions of protons with neutral hydrogen of the density $n\rs{H}$ \citep[e.g.][]{Kulsrud-Cesarsky-1971}. It is taken $\Gamma\rs{L}\propto n\rs{H}=0$ assuming that no neutral survive after passage across the shock front. There is no commonly accepted approach to the description of the non-linear damping \citep[see e.g.][and references therein]{Ptuskin-Zirak-2003}. 
We take the expression for the rate of non-linear Landau damping from \citet{Ptuskin-Zirak-2003}, namely from their Eq.~(12):
%\footnote{Note a misprint in Eq.~(12) of \citet{Ptuskin-Zirak-2003}: there should be $A^2$ instead of A.}:
\begin{equation}
 \Gamma\rs{NL}(k)W(k)=c\rs{K} v\rs{A} k\ \frac{E\rs{w}(>k)}{B^2/4\pi}\ W(k)
 \label{polariz:NLdampk}
\end{equation}
where $c\rs{K}=(2C\rs{K})^{-3/2}$, $C\rs{K}\approx3.6$ being the quantity introduced by \citet{Ptuskin-Zirak-2003}, and $E\rs{w}(>k)$ the energy density of waves with wave numbers larger than $k$. Note, that there is no explicit obliquity dependence in this term because the waves interact with themselves and this process does not depend on the shock obliquity.
%It is written per unit bandwidth of waves.
%The same integration \citep{amato2006} which leads to (\ref{polariz:eq3}) is now adopted:
Eq.~(\ref{polariz:NLdampk}) can be integrated over $k$, giving:
\begin{equation}
 \int%_{k\rs{min}}^{k\rs{max}}
 \Gamma\rs{NL}(k)W(k)dk=\frac{c\rs{K}v\rs{A}}{B^2/4\pi}\int%_{k\rs{min}}^{k\rs{max}}
 kE\rs{w}(>k)W(k)dk.
\end{equation}
Multiplying this expression by unity 
\begin{equation}
 1\equiv E\rs{w} \left(\int W(k)dk\right)^{-1},
\end{equation}
and noting that the average $\left\langle kE\rs{w}(>k)\right\rangle$ is dominated by the lowest wavenumber in the spectrum ($k\rs{min}$): $\left\langle kE\rs{w}(>k)\right\rangle\simeq k\rs{min}E\rs{w}$ 
% \citep[][]{Morlino-Caprioli-2012}
where $E\rs{w}(>k\rs{min})=E\rs{w}$, we have the damping term in Eq.~(\ref{polariz:eq1}) of the form:
\begin{equation}
 \Gamma\rs{NL}P\rs{w} =\frac{2c\rs{K}v\rs{A}}{B^2/8\pi}k\rs{min}P\rs{w}^2.
 \label{polariz:eq6c}
\end{equation}
Thus, Eq.~(\ref{polariz:eq1}) is now
\begin{eqnarray}
 &&\!\!\!\!\!\!\!\!\frac{\partial P\rs{w}}{\partial t}+u\frac{\partial P\rs{w}}{\partial r}+P\rs{w}\frac{3}{2}\frac{\partial u}{\partial r}=
 \nonumber\\
 &&\qquad\qquad \frac{v\rs{A}\cos\Theta}{2}\frac{\partial P\rs{c}}{\partial r}
 -\frac{c\rs{K}v\rs{A}}{B^2/8\pi}k\rs{min}P\rs{w}^2
\label{polariz:eq5a}
\end{eqnarray}
%\begin{equation}
% \frac{\partial P\rs{w}}{\partial t}+u\frac{\partial P\rs{w}}{\partial r}+P\rs{w}\frac{3}{2}\frac{\partial u}{\partial r}=
% \frac{v\rs{A}\cos(\Theta)}{2}\frac{\partial P\rs{c}}{\partial r}
% -\frac{c\rs{K}v\rs{A}\cos\Theta}{B^2/8\pi}k\rs{min}P\rs{w}^2
%\label{polariz:eq5a}
%\end{equation}

In order to follow more easily the evolution of waves in a given fluid element, we use the Lagrangian coordinate $a$ and the relations
\begin{equation}
 \left(\frac{\partial}{\partial t}+u\frac{\partial}{\partial r}\right)\rs{E}\!=\left(\frac{d}{dt}\right)\rs{L}\!,
\quad
%\end{equation}
%\begin{equation}
 \left(\frac{\partial}{\partial r}\right)\rs{E}\!=\frac{\rho(a)r(a)^2}{\rho\rs{o}(a)a^2}\left(\frac{\partial}{\partial a}\right)\rs{L}\!,
\end{equation}
where `E' and `L' denote Eulerian and Lagrangian derivatives respectively, and the latter relation directly follows from mass conservation ($\rho r^2dr=\rho\rs{o}a^2da$.) In terms of $a$, Eq.~(\ref{polariz:eq5a}) %and (\ref{polariz:eq5b}) 
has the form
\begin{equation}
 \frac{dP\rs{w}(a,t)}{dt}+q_1(a,t) P\rs{w}(a,t)+q_2(a,t) P\rs{w}(a,t)^2=q_0(a,t), %q->q_0, p->q_1
 \label{polariz:eq7}
\end{equation}
with
\begin{equation}
%q_0=\frac{B\cos(\Theta)\rho^{1/2} r^2(1-\alpha\rs{w})}{4\pi^{1/2} \rho\rs{o}a^2V}\frac{\partial P\rs{c}}{\partial a}.
 q_0=\frac{v\rs{A}\cos\Theta\ \rho r^2}{2 \rho\rs{o}a^2}\frac{\partial P\rs{c}}{\partial a},
 \label{polariz:eq7q}
\end{equation}
\begin{equation}
 q_1=\frac{3\rho r^2}{2\rho\rs{o}a^2}\frac{\partial u}{\partial a},
 \label{polariz:eq7p}
\end{equation}
\begin{equation}
 q_2=\frac{c\rs{K}v\rs{A}\ k\rs{min}}{B^2/8\pi}.
 %=\frac{0.7e\cos\Theta}{cp\rs{max}\rho^{1/2}V};
 \label{polariz:eq7q2}
\end{equation}
%in the last equality we used substitutions $v\rs{A}=B/\sqrt{4\pi \rho}$ and $k\rs{min}=1/r\rs{L}(p\rs{max})$.

Eq.~(\ref{polariz:eq7}) is the Riccati differential equation and may be solved numerically. Solutions of Eq.~(\ref{polariz:eq7}) are considered in detail in Sect.~\ref{polariz:sol-main-eq}. In the rest of this section, we describe how to model different properties of waves which are needed to solve this equation.

The solution $P\rs{w}(a,t)$, with known $r(a,t)$, $\rho(a,t)$, $u(a,t)$, $B(a,t)$, $P\rs{c}(a,t)$ describes the wave pressure. The evolution of $\delta B$ downstream is given by 
\begin{equation}
 \delta B(a,t)=\sqrt{8\pi P\rs{w}(a,t)}.
 \label{polariz:eq9}
\end{equation}
This dependence on $a$ may be converted to the spatial dependence on $r$ with the use of the relation $r(a,t)$ which is known explicitly from the hydrodynamic solution.

\subsection{Shock compression factor for waves}

Let us define the `shock compression factor' for the wave pressure as the ratio of pressures immediately downstream $P\rs{ws}$ and upstream $P\rs{wo}$:
\begin{equation}
 \sigma\rs{Pw}=P\rs{ws}/P\rs{wo}. 
\end{equation}
This ratio may be calculated from the stationary form ($\partial P\rs{w}/\partial t=0$) of Eq.~(\ref{polariz:eq1b}) written as
\begin{equation}
 \frac{1}{P\rs{w}}\frac{\partial P\rs{w}}{\partial r}+\frac{3}{2}\frac{1}{u}\frac{\partial u}{\partial r}=
 \frac{1}{2}\left(\sigma\rs{w}-\Gamma\rs{w}\right)\frac{dt}{dr}.
\label{polariz:eq5b}
\end{equation}
%where $\tau\rs{\sigma}=\sigma\rs{w}^{-1}$ and $\tau\rs{\Gamma}=\Gamma\rs{w}^{-1}$ are the time-scales for the wave growth and damping. 
By integrating this equation across the shock, we get:
\begin{equation}
 \ln\frac{P\rs{ws}}{P\rs{wo}}+\frac{3}{2}\ln\frac{u\rs{s}}{u\rs{o}}=
 \frac{\Delta t}{4\tau\rs{\sigma s}}+\frac{\Delta t}{4\tau\rs{\sigma o}}
 -\frac{\Delta t}{4\tau\rs{\Gamma s}}-\frac{\Delta t}{4\tau\rs{\Gamma o}},
\label{polariz:eq5c}
\end{equation}
where $\tau\rs{\sigma}=\sigma\rs{w}^{-1}$ and $\tau\rs{\Gamma}=\Gamma\rs{w}^{-1}$ are the time-scales for the wave growth and damping.
In order to come to this equation, we estimated the functions $\sigma\rs{w}$ and $\Gamma\rs{w}$ with their average: $(\sigma\rs{ws}+\sigma\rs{wo})/2$ and $(\Gamma\rs{ws}+\Gamma\rs{wo})/2$.

If the time-scale $\Delta t$ for the shock to cross the fluid element is much smaller than all $\tau\rs{\sigma s}$, $\tau\rs{\sigma o}$, $\tau\rs{\Gamma s}$ and $\tau\rs{\Gamma o}$ then the right side in (\ref{polariz:eq5c}) vanishes and then 
\begin{equation}
 \sigma\rs{Pw}=\sigma^{3/2}
 \label{polariz:sigmaPw}
\end{equation}
where we have used the relation $u\rs{o}=\sigma u\rs{s}$.
An important result is that, since $\sigma$ is constant for strong shocks, the factor $\sigma\rs{Pw}$ neither depends on time nor on the shock obliquity. The shock jump in $\delta B$ is then simply 
\begin{equation}
 \frac{\delta B\rs{s}}{\delta B\rs{o}}=\left(\frac{P\rs{ws}}{P\rs{wo}}\right)^{1/2}=\sigma^{3/4}. 
\end{equation}

\subsection{Model for the wave pressure on the shock $P\rs{ws}$}

The energy density of waves is assumed to reach the saturation level before the shock front. Its value $P\rs{wo}$ is given by the stationary form \citep{amato2006} of Eq.~(\ref{polariz:eq1b}) written in the upstream region:
\begin{equation}
 -2u\frac{\partial P\rs{w}}{\partial r}=\sigma\rs{w}P\rs{w}-\Gamma\rs{w}P\rs{w}.
\label{polariz:eq0NL}
\end{equation}

Let us introduce two parameters
\begin{equation}
 \xi=\left(\frac{\sigma\rs{w}P\rs{w}}{\Gamma\rs{w}P\rs{w}}\right)^{1/2}
     \frac{P\rs{w}}{P\rs{co}},
 \label{polariz:xidef}
\end{equation}
\begin{equation}
 \lambda=\frac{uP\rs{w}}{\left(\sigma\rs{w}P\rs{w}\cdot\Gamma\rs{w}P\rs{w}\right)^{1/2}}.
\end{equation}
Note from Eqs.~(\ref{polariz:eq3}) and (\ref{polariz:eq6c}) that $\xi$ and $\lambda$ are independent of $P\rs{w}$. Eq.~(\ref{polariz:eq0NL}) becomes, after multiplication by $\lambda\xi/(uP\rs{co})$,
\begin{equation}
 -2\lambda\xi\frac{\partial}{\partial r}\left(\frac{P\rs{w}}{P\rs{co}}\right)=
 \xi^2-\left(\frac{P\rs{w}}{P\rs{co}}\right)^2.
 \label{polariz:eq0bNL}
\end{equation}
The solution of (\ref{polariz:eq0bNL}) may be derived by separation of variables, assuming $\xi=\xi\rs{o}$ and $\lambda=\lambda\rs{o}$ constant on the scale of the precursor, i.e. from the shock to distance $x\rs{o}$ upstream of the shock. Thus, immediately upstream
\begin{equation}
 P\rs{wo}=P\rs{co}\xi\rs{o}
 \frac{\exp\left(\Omega\rs{o}\right)-1}
 {\exp\left(\Omega\rs{o}\right)+1}
 \label{polariz:Pwo0NL}
\end{equation}
where $\Omega\rs{o}=x\rs{o}/\lambda\rs{o}$. 
The CR and wave pressures are set to zero beyond the distance $x\rs{o}$ because our model does not consider seed CRs and waves (note that $0\leq P\rs{wo}/(P\rs{co}\xi\rs{o})\leq 1$). 

In order to see the meaning of the parameters $\xi\rs{o}$ and $\Omega\rs{o}$, we use Eqs.~(\ref{polariz:eq3}) and (\ref{polariz:eq6c}), written for the upstream region 
and use some simplifications. For instance, the minimum wavevector $\mathbf{k}\rs{min}$ (whose modulus appears in Eq.~\ref{polariz:eq6c}) is set to $\mathbf{k}\rs{o}\simeq \hat b\rs{o} r\rs{Lo}^{-1}$ where 
$r\rs{Lo}=p\rs{max}c/(eB\rs{o})$ is the Larmor radius of protons with maximum momentum in the field $B\rs{o}$. The value of the CR pressure gradient in the shock precursor may be estimated as $\partial P\rs{c}/\partial r\simeq P\rs{co}/x\rs{o}$ where $x\rs{o}$
%$x\rs{o}\simeq D(p\rs{max})/u\rs{o}$ \citep{amato2005} 
is the distance the same protons are able to diffuse off the shock (the size of the CR precursor); it is directed opposite to $\hat n\rs{o}$. The unit vectors $\hat b\rs{o}$ and $\hat n\rs{o}$ are in the directions of the upstream MF and of the normal to the shock respectively; their dot product is $\hat b\rs{o}\cdot \hat n\rs{o} = \cos\Theta\rs{o}$. 
%$D$ the diffusion coefficient. 
The distance $x\rs{o}\simeq D(p\rs{max})/u\rs{o}$ \citep{amato2005} where $D(p\rs{max})=r\rs{Lo}c/3$ is the Bohm diffusion coefficient. 
So, 
\begin{equation}
 \xi\rs{o}\simeq\left(\frac{\cos\Theta\rs{o}}{2c\rs{K}} \frac{3u\rs{o}}{c}%\frac{r\rs{Lo}}{x\rs{o}}
 \frac{P\rs{Bo}}{P\rs{co}}\right)^{1/2},
 \label{polariz:defxi}
\end{equation}
\begin{equation}
 \Omega\rs{o}\simeq\frac{v\rs{Ao}\cos\Theta\rs{o}}{u\rs{o}\xi\rs{o}}=\frac{\cos\Theta\rs{o}}{M\rs{Ao}\xi\rs{o}}
 \label{polariz:defOmega}
\end{equation}
where $2c\rs{K}\simeq 0.1$, and $M\rs{Ao}={u\rs{o}}/{v\rs{Ao}}$ is the Alfv\'en Mach number of the shock. 

Now, the post-shock pressure of waves is given by
%solution (\ref{polariz:Pwo0NL}) may be rewritten as
%\begin{equation}
% P\rs{wo}\simeq P\rs{co}\xi\rs{o}
%% \frac{\exp\left({v\rs{Ao}\cos\Theta\rs{o}}/{(\xi V)}\right)-1}
%% {\exp\left({v\rs{Ao}\cos\Theta\rs{o}}/{(\xi V)}\right)+1}.
% \frac{\exp\left({\cos\Theta\rs{o}}/{(M\rs{Ao}\xi\rs{o})}\right)-1}
% {\exp\left({\cos\Theta\rs{o}}/{(M\rs{Ao}\xi\rs{o})}\right)+1}
% \label{polariz:PwoNL}
%\end{equation}
%or, in different notation,
\begin{equation}
 P\rs{ws}\simeq  \sigma\rs{Pw}P\rs{co}\xi\rs{o}
 \frac{\exp\left(\Omega\rs{o}\right)-1}
 {\exp\left(\Omega\rs{o}\right)+1}.
 \label{polariz:PwoNLb}
\end{equation}
%The post-shock energy density of waves is given by $P\rs{ws}=\sigma\rs{Pw}P\rs{wo}$. 

%--------------------------------------------------------------
\begin{figure}
 \centering
 \includegraphics[width=8truecm]{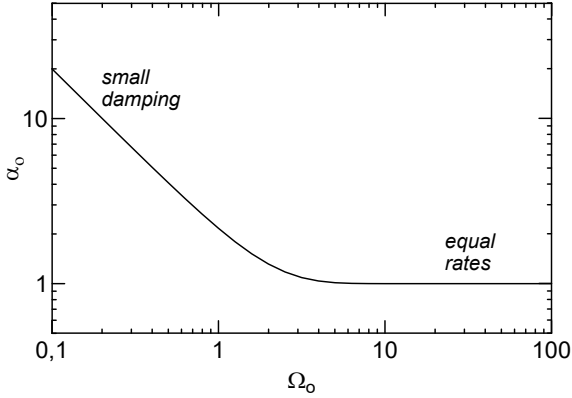}
 \caption{Ratio of the wave growth to damping rates $\alpha\rs{o}$ 
 as a function of $\Omega\rs{o}$.
               }
 \label{polariz:fig_alphalim}
\end{figure}
%--------------------------------------------------------------

\subsubsection{Limit cases of the wave evolution}
\label{polariz:subsectgenAB}

Let us introduce the variable $\alpha$ as the ratio of the wave growth and damping rates 
\begin{equation}
 \alpha=\frac{\sigma\rs{w}P\rs{w}}{\Gamma\rs{w}P\rs{w}}.
\end{equation}
Following the definition (\ref{polariz:xidef}), immediately before the shock 
\begin{equation}
 \xi\rs{o}=\alpha\rs{o}^{1/2}P\rs{wo}/P\rs{co}.
\label{polariz:xioalphao}
\end{equation}
Comparing this relation with the solution (\ref{polariz:Pwo0NL}), we see that the ratio between the two rates is related to the ratio of the exponents:
\begin{equation}
 \alpha\rs{o}=\left(\frac{\exp\left(\Omega\rs{o}\right)+1}
 {\exp\left(\Omega\rs{o}\right)-1}\right)^2.
 \label{polariz:alphadef}
\end{equation}

The plot $\alpha\rs{o}(\Omega\rs{o})$ is shown on Fig.~\ref{polariz:fig_alphalim}. The two limiting regimes of the wave evolution are evident from this plot: the limit A of the small damping rate at the shock ($\alpha\rs{o}=(2/\Omega\rs{o})^2$ is larger than unity, $\Omega\rs{o}\ll 2$) and the limit B of the same rates at the shock ($\alpha\rs{o}$ is unity, $\Omega\rs{o}\gg 2$). The maximum damping rate may not be larger than the growth one, as expected (we do not assume a seed turbulence). 

Fig.~\ref{polariz:fig_alphalim} demonstrates that the transition between the two limiting cases is in a rather narrow range. Namely, the first one may be used already for $\Omega<1$ and the second one for $\Omega>4$. 

It is apparent from the figure that the ratio of the growth to damping rates increases with obliquity, for the same $\xi\rs{o}$ and $M\rs{Ao}$, because $\Omega\rs{o}$ decreases with $\Theta\rs{o}$, Eq.~(\ref{polariz:defOmega}). 

Eq.~(\ref{polariz:PwoNLb}) gives the possibility to calculate the wave pressure at the present time, on the shock The wave pressure on the shock evolves as it follows from Eqs.~(\ref{polariz:PwoNLb}) and (\ref{polariz:alphadef}) and the expression for this evolution may be found as a ratio:
\begin{equation}
 \frac{P\rs{wsi}}{P\rs{ws}}=
 \frac{P\rs{csi}}{P\rs{cs}}
 \frac{\xi\rs{oi}}{\xi\rs{o}}
 \frac{\alpha\rs{o}(\Omega\rs{o})^{1/2}}{\alpha\rs{o}(\Omega\rs{oi})^{1/2}}
 \label{polariz:Pwsfin-gencase}
\end{equation}
where the index `i' refers to the time $t\rs{i}$, i.e. $P\rs{wsi}$ and $P\rs{csi}$ are the the wave and CR pressures in the element $a$ when it was shocked.
This formula is general; its limits (which are useful for interpretation of results) are considered in the next two subsections.

%It could be useful to relate parameters written above to the ratio of the turbulent component to the ordered component of MF at the shock. 
%In fact, we have from the definition $P\rs{ws}/P\rs{Bs}=(\delta B\rs{s}/B\rs{s})^2$, (\ref{polariz:defxi}) and 
%(\ref{polariz:PwoNLb}) that
%\begin{equation}
% \left(\frac{\delta B}{B}\right)\rs{s}^2=
%\frac{\sigma\rs{Pw}^{1/2}}{\sigma\rs{B}}
% \left(\frac{3VP\rs{co}}{2c\rs{K}cP\rs{Bo}\alpha\rs{o}}\right)^{1/4}.
% \label{polariz:Pws-deltaBB}
%\end{equation}

\subsubsection{Limit case A: small wave-damping at the shock}
\label{polariz:subsectlimA}

If the damping rate in the shock precursor is negligible compared to the growth rate (limit A), then Eq.~(\ref{polariz:eq0NL}) becomes 
\begin{equation}
 \frac{\partial P\rs{w}}{\partial r}=\frac{v\rs{Ao}\cos\Theta\rs{o}}{2u\rs{o}}\frac{\partial P\rs{c}}{\partial r}
 \label{polariz:Pwo0}
\end{equation}
with the obvious solution 
\begin{equation}
 P\rs{wo}=P\rs{co}v\rs{Ao}\cos(\Theta\rs{o})/(2u\rs{o}).
 \label{polariz:Pwo}
\end{equation}
The same expression may be obtained from the more general formula (\ref{polariz:PwoNLb}). Namely, if $\Gamma\rs{w}P\rs{w}\ll\sigma\rs{w}P\rs{w}$ then $\Omega\rs{o}\rightarrow 0$. Decomposition of the exponential terms with small arguments in (\ref{polariz:PwoNLb}) into series and use of the first-order term and Eq.~(\ref{polariz:defOmega}) result in (\ref{polariz:Pwo}).\footnote{This way to derive Eq.~(\ref{polariz:Pwo}) proves also the correctness of Eq.~(\ref{polariz:defOmega}).}

The CR pressure is smooth across the shock: $P\rs{co}=P\rs{cs}$. 
The post-shock pressure in form of waves is therefore 
\begin{equation}
 P\rs{ws}=\frac{\sigma\rs{Pw}P\rs{cs}v\rs{Ao}\cos\Theta\rs{o}}{2u\rs{o}}
 \label{polariz:Pws}
\end{equation}
in this limit.

In a non-uniform medium with a density profile $\rho\rs{o}(R)$, the velocity $V$ of the strong adiabatic shock is well approximated by $V(R)\propto \left(\rho\rs{o}(R) R^3\right)^{-1/2}$ \citep[][this formula is exact in the case of a constant density]{Hn-Pet-99}. Numerically, $u\rs{o}=V$. 
In the general case of an ambient medium with non-uniform distributions of density and MF, the variation of the post-shock value of the wave pressure in time follows from (\ref{polariz:Pws}). Namely, the evolution of the wave pressure on the shock may be found as the ratio 
\begin{equation}
 \frac{P\rs{wsi}}{P\rs{ws}}=
 \frac{P\rs{csi}}{P\rs{cs}}
 \left(\frac{a}{R}\right)^{3/2}
 \left\{
 \frac{B\rs{o}(a)\cos\Theta\rs{o}(a)}{B\rs{o}(R)\cos\Theta\rs{o}(R)}
 \right\}
 \label{polariz:Pwsfin}
\end{equation}
The term in braces $\{...\}=1$ for a uniform ISMF.

\subsubsection{Limit case B: equal rates at the shock}
\label{polariz:subsectlimB}

If $\Gamma\rs{w}P\rs{w}=\sigma\rs{w}P\rs{w}$ at the shock (limit B) then $\Omega\rs{o}$ is large. Unity may be neglected comparing to the exponential term both in the nominator and denominator in (\ref{polariz:PwoNLb}), i.e. $\alpha\rs{o}=1$ and the expression for pressure of waves comes simply from (\ref{polariz:xioalphao}):  
\begin{equation}
 P\rs{wo}= P\rs{co}\xi\rs{o}.
 \label{polariz:PwoNLlimB}
\end{equation}
Therefore, for a general case of non-uniform ISM and ISMF,
\begin{equation}
\begin{array}{ll}
 \displaystyle
 \frac{P\rs{wsi}}{P\rs{ws}}&\displaystyle =
 \left(\frac{P\rs{csi}}{P\rs{cs}}\right)^{1/2}
 \left(\frac{a}{R}\right)^{-3/4}
 \\ &\displaystyle\times
 \left\{
 \frac{B\rs{o}(a)\cos\Theta\rs{o}(a)^{1/2}}{B\rs{o}(R)\cos\Theta\rs{o}(R)^{1/2}}
 \frac{\rho\rs{o}(a)^{-1/4}}{\rho\rs{o}(R)^{-1/4}}
 \right\}
\end{array}
\label{polariz:Pwsfin-limB}
\end{equation}
The term in braces $\{...\}=1$ for ISM with uniform distributions of density and MF.
Equation (\ref{polariz:Pwsfin-limB}) is derived as a ratio $P\rs{wsi}/P\rs{ws}$ directly from (\ref{polariz:PwoNLlimB}) using expression (\ref{polariz:defxi}) and the approximation $V(R)\propto \left(\rho\rs{o}(R) R^3\right)^{-1/2}$.

\subsubsection{Obliquity dependence of $P\rs{ws}$}

%Sect.~\ref{polariz:subsectgenAB} presents a formula to calculate the evolution of the wave pressure at the shock; 
%Sects.~\ref{polariz:subsectlimA}-\ref{polariz:subsectlimB} considers two limits of this formula. 
In order to simulate polarization images, we need also to know how does the wave pressure vary with obliquity. 
We will use Eq.~(\ref{polariz:PwoNLb}) for this purpose, and have therefore to find the obliquity dependences for all the quantities there. 

The compression factor $\sigma\rs{Pw}$ is constant, Eq.(\ref{polariz:sigmaPw}).

The obliquity dependence of $\xi\rs{o}$ comes from Eq.~(\ref{polariz:defxi}). In order to find it, we write  
the ratio of this parameter for a parallel shock (marked with the additional index $\|$) and for the shock with a generic obliquity angle (without additional index). 
Then we have, for a general case of non-uniform ISM and ISMF, that 
\begin{equation}
 \xi\rs{o}\simeq\xi\rs{o\|} \cos\Theta\rs{o}^{1/2}
 \frac{P\rs{cs\|}^{1/2}}{P\rs{cs}^{1/2}}
 \left\{\frac{B\rs{o}\rho\rs{o\|}^{1/4}R\rs{\|}^{3/4}}{B\rs{o\|}\rho\rs{o}^{1/4}R^{3/4}}\right\}.
 \label{polariz:xiobliq}
\end{equation}
The term in braces is equal to unity for uniform ambient density and MF.
The obliquity variation of $\Omega\rs{o}$ is given by definition (\ref{polariz:defOmega}) and (\ref{polariz:xiobliq}): 
\begin{equation}
 \Omega\rs{o}=\Omega\rs{o\|}\cos\Theta\rs{o}^{1/2}
 \frac{P\rs{cs}^{1/2}}{P\rs{cs\|}^{1/2}}
 \left\{\frac{\rho\rs{o}^{1/4}R^{9/4}}{\rho\rs{o\|}^{1/4}R\rs{\|}^{9/4}}\right\} .
 \label{polariz:Omega-obliq}
\end{equation}
In deriving Eqs.~(\ref{polariz:xiobliq})-(\ref{polariz:Omega-obliq}), the approximate formula $V(R)\propto \left(\rho\rs{o}(R) R^3\right)^{-1/2}$ was used.

Now, we know how $\xi\rs{o}$ and $\Omega\rs{o}$ vary with obliquity. Thus, Eq.~(\ref{polariz:PwoNLb}) gives us the ratio
\begin{equation}
 \frac{P\rs{ws}}{P\rs{ws\|}}=
 \frac{P\rs{cs}}{P\rs{cs\|}}
 \frac{\xi\rs{o}}{\xi\rs{o\|}}
 \frac{\alpha\rs{o}(\Omega\rs{o\|})^{1/2}}{\alpha\rs{o}(\Omega\rs{o})^{1/2}}
\end{equation}
which represents the obliquity dependence of the wave pressure.
This reduces, for the Sedov shocks, to 
\begin{equation}
 \frac{P\rs{ws}}{P\rs{ws\|}}=
 \frac{P\rs{cs}}{P\rs{cs\|}}\cos\Theta\rs{o}
 \label{polariz:Pwsfin-caseA}
\end{equation}
in the limit A and to 
\begin{equation}
 \frac{P\rs{ws}}{P\rs{ws\|}}=
 \left(\frac{P\rs{cs}}{P\rs{cs\|}}\right)^{3/2}\cos\Theta\rs{o}^{1/2}
 \label{polariz:Pwsfin-caseB}
\end{equation}
in the limit B.

\subsection{Evolution of the CR pressure $P\rs{c}$}
\label{polariz:sectPc}

For a given momentum distribution $f(p)$ of CRs, the CR pressure is
\begin{equation}
 P\rs{c}=\frac{4\pi}{3}\int\limits p^3v(p)f(p)dp.
 \label{palariz:Pcdef}
\end{equation}

In order to simplify the calculations of the evolution of $\delta B$, we approximate $P\rs{c}$ assuming an isotropic test-particle CR distribution $N(p)=4\pi p^2f(p)$ which is represented by a power law $N(p)=Kp^{-s}$, with $s$ close to $2$, in the range between the minimum and maximum momenta of CRs $p\rs{min}$ and $p\rs{max}$. The CR pressure in this case is
\begin{equation}
 P\rs{c}(a,t)=\frac{1}{3}K(a,t)\int\limits_{p\rs{min}(a,t)}^{p\rs{max}(a,t)} p^{1-s}v(p)dp.
 \label{palariz:Pc1}
\end{equation}
Since an approximate treatment of $P\rs{c}(r)$ is sufficient to our purposes, in this subsection, we fix $s=2$, a choice that minimizes the dependence of the CR pressure on $p\rs{max}$, while describing reasonably well also the cases with $s$ slightly different from $2$. 

Then, in the regime $p\rs{min}\ll mc\ll p\rs{max}$, being $v\simeq c$ for $p\geq mc$, we finally derive 
\begin{equation}
 P\rs{c}(a,t)=\frac{cK\rs{s}(t)}{3}\frac{K(a,t)}{K\rs{s}(t)}\ln\left(\frac{p\rs{max}(a,t)}{mc}\right).
 \label{palariz:Pc2}
\end{equation}
In all reasonable cases the logarithmic term can be approximated with $\ln\left({p\rs{max,s}}/{mc}\right)$, where $p\rs{max,s}\geq 10^3mc$.

The normalization $K\rs{s}$, related to the injection efficiency $\eta$ by
\begin{equation}
 \int_{p\rs{min,s}}^{p\rs{max,s}}K\rs{s}p^{-2}dp=\eta n\rs{s},
\end{equation}
evaluates
\begin{equation}
 K\rs{s}=\eta n\rs{s} p\rs{min,s}.
 \label{polariz:Kseta}
\end{equation} 
At the shock ($a=R$), the expression for the CR pressure, Eq.~(\ref{palariz:Pc2}) with (\ref{polariz:Kseta}), coincides with Eq.~(30) in \citet{amato2009}. 

The minimum momentum may be estimated by its relation to the thermal momentum. We assume that the distribution of the thermal particles downstream may be approximated by the Maxwellian 
\begin{equation}
 f\rs{M}(p)=\frac{4}{\sqrt{\pi}}y^2\exp(-y^2)
\end{equation}
where $y=p/p\rs{th}$ with $p\rs{th}=\sqrt{2mkT\rs{s}}$ is the thermal momentum and 
\begin{equation}
 T\rs{s}=\frac{2(\gamma-1)}{(\gamma+1)^2}\frac{\mu\rs{o} m}{k}V^2 
\end{equation}
is the shock temperature, $\mu\rs{o}$ the mean particle mass in units of the proton mass. 
Further, we assume that all thermal particles with momenta $p>p\rs{min}$ are to be accelerated. Thus, the injection efficiency is
\begin{equation}
\eta=\int_{y\rs{min}}^{\infty}f\rs{M}(y)dy.
\end{equation}
In general, in order to find the minimum momentum $p\rs{min,s}=y\rs{min}p\rs{th}$, one has to solve this equation to find $y\rs{min}$ for a given value of $\eta$. 
This formula gives, in particular, $y\rs{min}=3.07$ for $\eta=3\E{-4}$ and $y\rs{min}=3.91$ for $\eta=10^{-6}$. 
Since $y\rs{min}$ varies quite slowly for a wide range of $\eta$, we set in our simulations $p\rs{min,s}= 3.4 p\rs{th}(T\rs{s})$. 

Taking all these pieces together, 
\begin{equation}
 P\rs{c}(a)\simeq\frac{2.3\eta\rho\rs{o}Vc}{\mu\rs{o}^{1/2}(\gamma-1)^{1/2}}
\frac{K(a)}{K\rs{s}}
\ln\left(\frac{p\rs{max,s}}{mc}\right)
\label{polariz:Pcfin}
\end{equation}
where all parameters are referred to the time $t$ and some of them depend on the shock obliquity. 

Let us parametrize the time dependence of the injection efficiency as $\eta(t)\propto V(t)^{-(b+1)}$ with a parameter $b$ (i.e. $K\rs{s}\propto V^{-b}$ because $p\rs{min}\propto V$). Then the ratio of CR pressures needed in (\ref{polariz:Pwsfin-gencase}) is
\begin{equation}
 \frac{P\rs{csi}}{P\rs{cs}}=\frac{\rho\rs{o}(a)}{\rho\rs{o}(R)}\left(\frac{V(a)}{V(R)}\right)^{-b}
 =\left(\frac{\rho\rs{o}(a)}{\rho\rs{o}(R)}\right)^{1+b/2}\left(\frac{a}{R}\right)^{3b/2}.
 \label{polariz:Pcsfin}
\end{equation}

The way to calculate $K(a)/K\rs{s}$ is described in \citet{Orlando-etal-2007,Orlando-etal-2011} for general case of nonuniform ISM and ISMF and in \citet{xmaps} for a Sedov shocks in ISM with uniform density and MF.
In particular, the evolution of the normalization $K$ behind the Sedov shock is self-similar: 
$K(a,t)/K\rs{s}(t)=\bar{K}(\bar a)$ where 
\begin{equation}
 \bar{K}(\bar a)=
 \bar{a}^{\ \!3b/2}\ 
 \bar{\rho}(\bar{a})^{(2+s)/3}
 \label{self-K}
\end{equation}
is time-independent, $\bar a=a/R$, $R$ the radius of the shock. 
The profile of the CR pressure, Eq.~(\ref{polariz:Pcfin}), is 
(to the level of approximation used for the logarithmic factor in Eq.~\ref{palariz:Pc2}) 
self-similar as well, i.e. it may be written in the form $P\rs{c}(a,t)=P\rs{cs}(t)\bar P\rs{c}(\bar a)$ where
\begin{equation}
 P\rs{cs}(t)=\frac{2.3\eta\rho\rs{o}Vc}{\mu\rs{o}^{1/2}(\gamma-1)^{1/2}}
\ln\left(\frac{p\rs{max,s}}{mc}\right),
 \label{polariz:Pcs}
\end{equation}
\begin{equation}
 \bar P\rs{c}(\bar a)=\bar K(\bar a). %\left(1+0.07\ln(\bar a^{-3q/2}\bar \rho^{1/3})\right); 
 \label{self-Pc}
\end{equation}
%in the latter expression, we put a number 0.07 as an approximation of the term $1/\ln\left({p\rs{max,s}}/{mc}\right)$ for $E\rs{max}=10^{4}\div 10^{6}\un{GeV}$.

We have just described the time evolution of the CR pressure, at the parallel shock. We also need the prescription for the whole SNR surface, i.e. at the oblique shocks as well. 
The obliquity dependence of $P\rs{cs}$ is due to the respective dependence of the injection efficiency and variations (if any) of the ambient density and the shock radius. It is given by the ratio
\begin{equation}
 \frac{P\rs{cs}}{P\rs{cs\|}}=
 \frac{\eta}{\eta\rs{\|}}
 \left\{\frac{\rho\rs{o}^{1/2}R\rs{\|}^{3/2}}{\rho\rs{o\|}^{1/2}R^{3/2}}\right\}
\end{equation}
derived from (\ref{polariz:Pcs}) where we neglected for simplicity the obliquity dependence of the logarithmic term and used the approximate formula $V(R)\propto \left(\rho\rs{o}(R) R^3\right)^{-1/2}$.

\subsection{Evolution of $k\rs{min}$}
\label{polariz:sect-kmin}

The function $q_2$, Eq.~(\ref{polariz:eq7q2}), contains $k\rs{min}$. 
In our model, the downstream evolution of $k\rs{min}$ is described as follows. Since $W(k)$ is a decreasing function, the damping is highest for waves with the smallest $k$, Eq.~(\ref{polariz:NLdampk}). Therefore, we assume that, in a given fluid element, at a given time, the only survived waves are those with the wavenumbers larger than $k\rs{min}$ which is able to interact resonantly with the Larmor radius of particles of the highest energy: $k\rs{min}\simeq 1/r\rs{L}(p\rs{max})=eB/(cp\rs{max})$. Then, we consider the evolution of $B$ and $p\rs{max}$ within this fluid element. 

In order to describe $p\rs{max}(a,t)$, we parametrize the time evolution of the maximum momentum which particles reach at the shock as $p\rs{max,s}(t)\propto V(t)^q$; then
\begin{equation}
 \frac{p\rs{max,s}(t\rs{i})}{p\rs{max,s}(t)}=\left(\frac{V(a)}{V(R)}\right)^q=\left(\frac{a}{R}\right)^{-3q/2}
 \label{polariz:pmaxatshock}
\end{equation}
Numerically, {for the Sedov phase, $q=1/3$} if $p\rs{max}$ is time-limited\footnote{{In this case, $p\rs{max}\propto V^2 t$ \citep{Lagage-Cesarsky-1983} and $V\propto t^{-3/5}$ for the Sedov shock; therefore, $p\rs{max}\propto V^{1/3}$.}}, $q=0$ if it is escape-limited and $q=1$ if it is limited by the radiative losses \citep[][{p.~379}]{Reyn-98}.
Downstream, even protons with the maximum momentum $p\rs{max}$ experience only adiabatic losses, i.e. $p\rs{max}(a)\propto E\rs{max}(a)\propto \rho(a)^{1/3}$ \citep[e.g.][]{Reyn-98}; the energy losses due to pion production is neglected assuming that the shock is propagating in medium with density less than $\sim 10^4\U{cm^{-3}}$ \citep[e.g.][]{ppmaps}. 
The evolution of $p\rs{max}$ downstream is therefore given by 
\begin{equation}
 \frac{p\rs{max}(a,t)}{p\rs{max,s}(t\rs{i})}=\left(\frac{\rho(a,t)}{\rho\rs{s}(t\rs{i})}\right)^{1/3}.
 \label{polariz:pmaxdown}
\end{equation}
Taking (\ref{polariz:pmaxatshock}) and (\ref{polariz:pmaxdown}) together,
\begin{equation}
 \frac{p\rs{max}(a,t)}{p\rs{max,s}(t)}=\left(\frac{a}{R}\right)^{-3q/2}\left(\frac{\rho(a,t)}{\rho\rs{s}(t)}\right)^{1/3}\left(\frac{\rho\rs{o}(a)}{\rho\rs{o}(R)}\right)^{-1/3}.
\end{equation}

It appears from the above formulae, that the evolution of the minimum wavenumber is self-similar for the Sedov shock:
\begin{equation}
 k\rs{min}(a,t)=k\rs{min,s}(t)\cdot\bar k\rs{min}(\bar a),
 \label{polariz:kminself}
\end{equation}
with 
\begin{equation}
 k\rs{min,s}(t)=\frac{eB\rs{s}}{p\rs{max,s}c},
\end{equation}
\begin{equation}
 \bar k\rs{min}=\bar B/\bar p\rs{max}=\bar B \bar a^{3q/2}\bar\rho^{\ \!-1/3}.
\end{equation}

%%%%%%%%%%%%%%%%%%%%%%%%%%%%%%%%%%%%%%%%%%%%%%%%%%%%%%%%%%%%%%%%%%%%%%%%%
\section{Solutions of equation for waves}
\label{polariz:sol-main-eq}

The evolution of the wave pressure $P\rs{w}$ within a fluid element $a$ after its passage through the shock is given by Eq.~(\ref{polariz:eq7}). In this section, we consider solutions of this equation. Sect.~\ref{polariz:sect:solA} deals with a particular case which allow for analytic expression. The formulae are presented for the general case and for a particular model of Sedov shocks in uniform medium and uniform MF. Numerical solutions are presented in Sect.~\ref{polariz:sect:solS}.

\subsection{Limit case A: small wave damping}
\label{polariz:sect:solA}

\subsubsection{Solution for a general case}

If the damping rate is negligible at the shock (i.e. $\Omega\rs{o}<1$) and everywhere downstream then the damping term in Eq.~(\ref{polariz:eq7}) may be set $q_2P\rs{w}^2=0$ and the solution of Eq.~(\ref{polariz:eq7}) is 
%given by Eq.~(\ref{polariz:eq8}), see Fig.~\ref{polariz:fig_deltaBb}.
\begin{equation}
 P\rs{w}(a,t)=\mu(a,t)\left(P\rs{ws}(t\rs{i})+\int_{t\rs{i}}^{t}\frac{q_0(a,t')}{\mu(a,t')}dt'\right), 
 \label{polariz:eq8a}
\end{equation}
\begin{equation}
 \mu(a,t)=\exp\left(-\int_{t\rs{i}}^{t}q_1(a,t')dt'\right).
 \label{polariz:eq8ab}
\end{equation}
In order to evaluate this solution numerically, one needs to know the time history of the whole SNR structure; in other words, one should have the MHD datacube for each time moment $t\rs{i}$.
It is more suitable for calculations to express the integrals in terms of $a$ rather than $t'$. With the use of
\begin{equation}
 dt'=\frac{dR(t')}{V(t')}=\frac{da}{V(a)},
 \label{polariz:eq8ccc}
\end{equation}
one has the spatial profile of the wave pressure
\begin{equation}
 P\rs{w}(a,t)=\mu(a,t)\left(P\rs{ws}\left[\frac{P\rs{wsi}}{P\rs{ws}}\right]+
  \int_{a}^{R(t)}\!\!\frac{q_0(a',t)}{\mu(a',t)V(a')} da'\right), 
 \label{polariz:eq8}
\end{equation}
\begin{equation}
 \mu(a,t)=\exp\left(-\int_{a}^{R(t)}\frac{q_1(a',t)}{V(a')}da'\right),
 \label{polariz:eq8b}
\end{equation}
$P\rs{wsi}\equiv P\rs{ws}(t\rs{i})$ is the value at time $t\rs{i}$, while other quantities are referred to time $t$. 
In this approach, it is enough to know the present-time SNR structure only. 
Having the present time value of $P\rs{ws}$ and the ratio $P\rs{wsi}/P\rs{ws}$ from Eq.~(\ref{polariz:Pwsfin}) (which, like the integrals in (\ref{polariz:eq8}) and (\ref{polariz:eq8b}), depends on the distribution of MHD parameters at the present time), we may calculate $P\rs{w}(a,t)$ numerically.

\subsubsection{Solution for a Sedov shock}
\label{polariz:sectsedlimA}

Let us consider this solution for a particular model, namely, for a Sedov shock in a uniform medium and uniform MF. 

For a Sedov SNR evolution in a medium with uniform density and MF Eq.~(\ref{polariz:eq8}) becomes 
\begin{equation}
 P\rs{w}(a,t)=P\rs{ws}(t)\big(\bar P\rs{wa}(\bar a) + \bar P\rs{wb}(\bar a) \big)
 \label{polariz:eq12}
\end{equation}
where
the profiles $P\rs{wa}(\bar a)$, $P\rs{wb}(\bar a)$ are time independent, $\bar a=a/R$. Namely,
\begin{equation}
 \bar P\rs{wa}(\bar a)=\bar a^{3(b+1)/2} \bar \mu(\bar a),
 \label{polariz:PwaSedov}
\end{equation}
\begin{equation}
 \bar P\rs{wb}(\bar a)=\frac{\sigma\rs{B}\bar \mu(\bar a)}{\sigma\cos(\Theta\rs{o})}
 \int^{1}_{\bar a}\frac{\bar B\bar\rho^{1/2}\bar r^2\cos(\Theta)}{\bar \mu \bar a'^{1/2}}
 \frac{d\bar P\rs{c}}{d\bar a'}d\bar a',
 \label{polariz:PwbSedov}
\end{equation}
\begin{equation}
 \bar \mu(\bar a)=\exp\left[-\frac{3}{2}\int^{1}_{\bar a}\frac{\bar\rho\bar r^2}{\bar a'^{1/2}}
 \frac{d\bar u}{d\bar a'}d\bar a'\right].
 \label{polariz:muSedov1}
\end{equation}
In order to derive Eq.~(\ref{polariz:PwaSedov}), we made use of Eqs.~(\ref{polariz:Pwsfin}), (\ref{polariz:Pcsfin}) and $u\rs{s}=V/\sigma$.
In derivation of (\ref{polariz:PwbSedov}), we made use of Eqs.~(\ref{polariz:Pws}), (\ref{polariz:sigmaPw}) and $V(R)/V(a)=\bar a^{3/2}$. %and $\cos(\Theta\rs{o})=\sigma\rs{B}\cos(\Theta\rs{s})$.

%--------------------------------------------------------------
\begin{figure}
 \centering
 \includegraphics[width=8.3truecm]{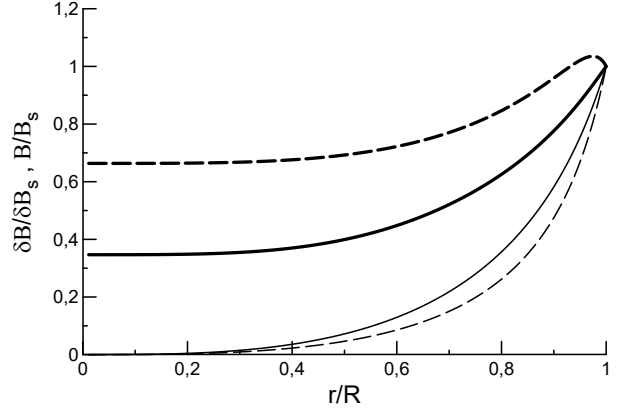}
 \caption{Normalized radial profiles of $\bar B(\bar r)$ (thin lines) and $\overline{\delta B}(\bar r)$ (thick lines) along SNR radius downstream of the parallel shock (solid lines) and perpendicular one (dashed lines). Limit of the small damping, $b=0$, $q=0$.
               }
 \label{polariz:fig_deltaBb}
\end{figure}
%--------------------------------------------------------------

Eqs.~(\ref{polariz:PwaSedov})-(\ref{polariz:muSedov1}) are written in a way to express the solution through the basic profiles of MHD parameters. It is clear from here that the downstream evolution of $P\rs{w}$ is self-similar. Eq.~(\ref{polariz:eq9}) shows that $\delta B(a)$ is  self-similar as well, $\delta B(a,t)=\delta B\rs{s}(t)\cdot \overline{\delta B}(\bar a)$, with 
\begin{equation}
 \delta B\rs{s}=\sqrt{8\pi P\rs{ws}}
\end{equation}
and $\overline{\delta B}=\bar P\rs{w}^{1/2}$, i.e.
\begin{equation}
 \overline{\delta B}=\left(\bar P\rs{wa}(\bar a)+\bar P\rs{wb}(\bar a)\right)^{1/2}.
\label{polariz:deltaBfin}
\end{equation}

The profiles $\overline{\delta B}(\bar r)$ are shown on Fig.~\ref{polariz:fig_deltaBb} in comparison with profiles of $\bar B(r)$. 
In the considered limit A, the turbulent MF has a flat distribution in the deep interior (no wave damping) while $B$ decreases. 
As a results, the ratio $\delta B/B$ increases toward the center of SNR.
The obliquity affects the normalized profiles $\overline{\delta B}(\bar r)$: the normalized strength of the turbulent MF is larger for a perpendicular shock. However, in the considered limit, the postshock value $\delta B\rs{s}\propto P\rs{ws}^{1/2}\propto \cos\Theta\rs{o}^{1/2}$ (Eq.~\ref{polariz:Pwsfin-caseA}), i.e. it zero for perpendicular shock.  
%(Fig.~\ref{polariz:fig_deltaBbb}).

\subsection{General solution for Sedov shock}
\label{polariz:sect:solS}

In a general case when both the wave growth and damping should be considered, the solution of Eq.~(\ref{polariz:eq7}) may be found numerically. 

Let us consider Sedov SNR evolution in medium with uniform density and MF and rewrite Eq.~(\ref{polariz:eq7}) in order to have an equation for $\bar P\rs{w}(\bar a)$. 
First, we substitute Eq.~(\ref{polariz:eq7}) with the self-similar form
\begin{equation}
 P\rs{w}(a,t)=P\rs{ws}(t)\bar P\rs{w}(\bar a), \qquad a=\bar a\cdot R(t).
\end{equation}
Then we use the same procedure as before, namely, transform from the variable $t$ to the variable $a$ with Eq.~(\ref{polariz:eq8ccc}). 
These steps yield the following equation for the normalized wave pressure $\bar P\rs{w}(\bar a)$: 
\begin{equation}
 \frac{d\bar P\rs{w}}{d\bar a}+{\bar q_1}\bar P\rs{w}+{\bar q_2}\bar P\rs{w}^2={\bar q_0} 
 \label{polariz:eq7bSedov}
\end{equation}
with 
\begin{eqnarray}
 {\bar q_0}&\!\!\!= \displaystyle\frac{q_0R}{V(a)P\rs{ws}} =\!\!\!&
 \frac{\sigma\rs{B}\cos\Theta}{\sigma\cos\Theta\rs{o}}
 \frac{\bar B \bar\rho^{1/2}\bar r^2}{\bar a^{1/2}}\frac{d \bar P\rs{c}}{d \bar a}
 \cdot \frac{\Omega\rs{o}\alpha\rs{o}^{1/2}}{2},\\
 {\bar q_1}&\!\!\!=\displaystyle \frac{q_1R}{V(a)} =\!\!\!&
 \frac{3\bar\rho\bar r^2}{2\bar a^{1/2}}\frac{d\bar u}{d\bar a},\\
 {\bar q_2}&\!\!\!=\displaystyle \frac{q_2RP\rs{ws}}{V(a)} =\!\!\!&
% \frac{\cos\Theta}{\cos\Theta\rs{o}}
 \frac{\bar a^{3(q+1)/2}}{\bar\rho^{5/6}}
 \frac{\sigma R}{x\rs{o}}
 \cdot \frac{\Omega\rs{o}\alpha\rs{o}^{-1/2}}{2}.
\end{eqnarray}

The factor ${\Omega\rs{o}\alpha\rs{o}^{-1/2}}{2^{-1}}=(\Omega\rs{o}/{2})^2$ in the limit of small wave damping. Therefore, $q_2\rightarrow 0$ in this limit because $\Omega\rs{o}\ll 2$. In addition, ${\Omega\rs{o}\alpha\rs{o}^{1/2}}{2^{-1}}=1$ and we recover results of Sect.~\ref{polariz:sectsedlimA}. 

In the limit of equal rates $\alpha\rs{o}^{1/2}=1$. It should also be $\bar q_2 \bar P\rs{w}^2=\bar q_0$ at the shock. This yields 
\begin{equation}
 \frac{R}{x\rs{o}}=\frac{1}{\sigma^2}\left[\frac{d \bar P\rs{c}}{d \bar a}\right]_{s},
 \label{polariz2:rxo}
\end{equation}
where we have used the relation $\cos\Theta\rs{o}=\sigma\rs{B}\cos\Theta\rs{s}$.
Though the expression (\ref{polariz2:rxo}) comes from a limit, we shall use it for any $\alpha\rs{o}$. Therefore, we adopt
\begin{equation}
 {\bar q_2}=
 %\frac{\sigma\rs{B}\cos\Theta}{\sigma\cos\Theta\rs{o}}
 \frac{\bar a^{3(q+1)/2}}{\sigma\bar\rho^{5/6}}
 \left[\frac{d \bar P\rs{c}}{d \bar a}\right]_{s}
 \cdot \frac{\Omega\rs{o}\alpha\rs{o}^{-1/2}}{2}.
\end{equation}
With (\ref{self-Pc}), (\ref{self-K}) and, for Sedov solutions \citep{Hn-Pet-96,Hn-Pet-99}
\begin{equation}
 \left[\frac{d \bar \rho}{d \bar a}\right]_{s}=\frac{5\gamma+13}{(\gamma+1)^2},
\end{equation}
we have
\begin{equation}
 \left[\frac{d \bar P\rs{c}}{d \bar a}\right]_{s}=
 \frac{3b}{2}+\frac{2+s}{3}\frac{5\gamma+13}{(\gamma+1)^2}.
\end{equation}
We use $s=2$ and $\gamma=5/3$ for plots and numerical simulations.

In order to have the {\it spatial profile} $\bar P\rs{w}(\bar a)$, we should start from the initial value of $P\rs{w}(a,t\rs{i})$ in the fluid element $a$ at time $t\rs{i}$ when it was shocked, i.e. from $P\rs{wsi}(a)$; its normalized version is $\bar P\rs{wsi}(\bar a\rs{ini})$. Then the Eq.~(\ref{polariz:eq7bSedov}) has to be used to evolve the wave pressure in this fluid element up to the time $t$, i.e. to find $P\rs{w}(a,t)$; its normalized version is $\bar P\rs{w}(1)$. Note that $\bar P\rs{w}(1)$ is different for different initial values $\bar P\rs{wsi}(\bar a)$. In other words, the solution $\bar P\rs{w}(\bar a)$, $\bar a\in [\bar a\rs{ini},1]$, of Eq.~(\ref{polariz:eq7bSedov}) with the initial value $\bar P\rs{wsi}(\bar a\rs{ini})$ does not represents the spatial distribution of the wave pressure for different $\bar a$ but the {\it time evolution} of $P\rs{w}$ within the element $a$ written in terms of time in an indirect way. %This is why $\bar P\rs{w}(1)$ gives the value of $\bar P\rs{w}$ in the element $\bar a$ at time $t$ in the solution of Eq.~(\ref{polariz:eq7bSedov}). 
Therefore, in order to have the {\it spatial variation} $\bar P\rs{w}(\bar a)$, one need to solve this equation many times with the initial conditions for different $\bar a\rs{ini}$. Different values of $\bar P\rs{w}(1)$ derived from each solution correspond to the wave pressure in respective fluid elements $\bar a\rs{ini}=a/R(t)$ at time $t$ and therefore represent such a spatial variation, $\bar P\rs{w}(\bar a\rs{ini})$. 

This could look like a complicate procedure. However, the direct solution of the equation (\ref{polariz:eq7}) is not simpler from the point of view of numerical realization. Namely, in our approach, we do not need to keep the time history of spatial distributions of all the parameters to make temporal integration. We just take one, the present-time spatial distribution of parameters and do the spatial integration instead of the temporal one (cf. Sect.~\ref{polariz:sect:solA}).

The initial condition, as derived from Eqs.~(\ref{polariz:Pwsfin-gencase}), (\ref{polariz:defxi}) and (\ref{polariz:Pcsfin}), is:
\begin{equation}
 \bar P\rs{wsi}\equiv \frac{P\rs{ws}(t\rs{i})}{P\rs{ws}(t)}=
 \bar a^{3(b-1)/4}
 \left[\frac{\alpha\rs{o}\left(\Omega\rs{o}\right)}{\alpha\rs{o}(\Omega\rs{o}\bar a^{3(b+3)/4})}\right]^{1/2},
 \label{polariz:Pwsi}
\end{equation}
where we substituted $\Omega\rs{oi}=\Omega\rs{o}\bar a^{3(b+3)/4}$ (as from Eqs.~\ref{polariz:defOmega}, \ref{polariz:defxi}, \ref{polariz:Pcsfin}) and assumed a uniform ambient medium density and MF. 
In the limit of small damping, this results in 
\begin{equation}
 \bar P\rs{wsi}=\bar a^{3(b+1)/2}
 \label{polariz:PwsiA}
\end{equation}
(cf. Eq.~\ref{polariz:Pwsfin}); while in the limit of equal rates we obtain (cf. Eq.~\ref{polariz:Pwsfin-limB})
\begin{equation}
 \bar P\rs{wsi}=\bar a^{3(b-1)/4}.
 \label{polariz:PwsiB}
\end{equation}

%--------------------------------------------------------------
\begin{figure}
 \centering
 \includegraphics[width=8.3truecm]{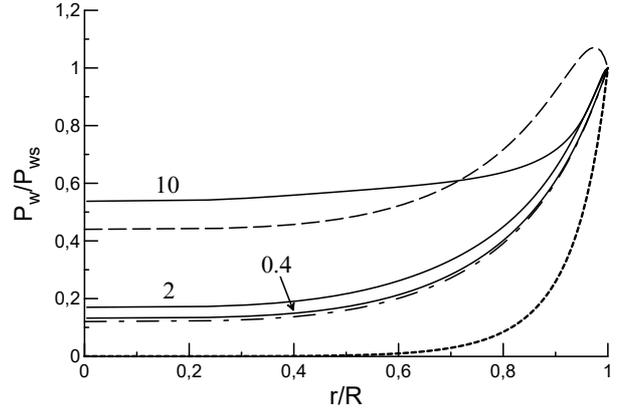}
 %\vspace{5truecm}
 \caption{Profiles $\bar P\rs{w}(\bar r)$ as from Eq.~(\ref{polariz:eq7bSedov}) for $\Omega\rs{o}=0.4,2,10$ (solid lines). The solution in the limit of negligible damping (Eq.~\ref{polariz:eq12}) is shown by the dot-dashed line. For the plot we have chosen $b=0$, $q=0$. The solid and dot-dashed lines refer to the parallel shock, while the long-dashed line to a  perpendicular shock ($\Omega\rs{o}=0$ for the perpendicular shock, Eq.~\ref{polariz:Omega-obliq}). {The profile $\bar P\rs{c}(\bar r)$,  Eq.~(\ref{self-Pc}), is shown by the dotted line for comparison.} 
               }
 \label{polariz:fig_deltaBgen1}
\end{figure}
%--------------------------------------------------------------

\begin{figure}
 \centering
 \includegraphics[width=8.3truecm]{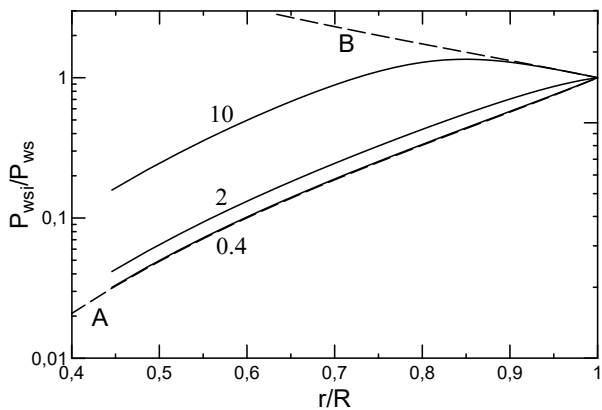}
 %\vspace{5truecm}
 \caption{Values of $\bar P\rs{wsi}$ (as from Eq.~\ref{polariz:Pwsi}) for the fluid element $\bar a$ at time $t\rs{i}$ when it was shocked plotted versus the present position $\bar r$ of this element. All cases are for a parallel shock with $b=0$; lines are labeled with the values of $\Omega\rs{o}$. The same for the limit A (Eq.~\ref{polariz:PwsiA}; dashed line marked by 'A') and for the limit B (Eq.~\ref{polariz:PwsiB}; dashed line marked by 'B'). 
               }
 \label{polariz:fig_deltaBgen4}
\end{figure}

Fig.~\ref{polariz:fig_deltaBgen1} shows the radial distribution of the wave pressure downstream of the Sedov shock for some values of $\Omega\rs{o}$. The larger the $\Omega\rs{o}$ the closer the damping rate to the rate of the wave growth at the shock (Fig.~\ref{polariz:fig_alphalim}). However, for larger $\Omega\rs{o}$, the values of $\bar P\rs{wsi}$ were higher at previous times (Fig.~\ref{polariz:fig_deltaBgen4}). Combination of these features results in the larger wave pressure $P\rs{w}$ in the interior of SNR for larger $\Omega\rs{o}$ (Fig.~\ref{polariz:fig_deltaBgen1}).

How to know which regime of the wave behavior (limit A or B) is more reliable for a certain parameter set? 
One can note from Eqs.~(\ref{polariz:defOmega}), (\ref{polariz:defxi}) and (\ref{polariz:Pcs}) that 
\begin{equation}
 \Omega\rs{o}\simeq 1.5 \eta^{1/2} \frac{c}{V} \cos\Theta\rs{o}^{1/2}.
 \label{polariz:OmegacVcos}
\end{equation}
This means that a typical SNR shock ($\eta^{1/2}c/V\sim 0.1\div 10$) could be either in the small damping limit or in the limit of the equal damping and growth rates. The quasi-perpendicular shocks are always in the regime of the small damping because of the dependence of $\Omega\rs{o}$ on the cosine of the obliquity angle.

The solution for the limit of the small damping, Eq.~(\ref{polariz:eq12}), is shown on Fig.~\ref{polariz:fig_deltaBgen1} by the {dot-}dashed line. The radial profiles for small $\Omega\rs{o}$ values are close to that approximate solution. The differences are evident deep downstream. The solution in the limit A is obtained under assumption that the condition $q_2P\rs{w}^2\ll q_0$ is valid everywhere inside the SNR. However, in the solution of the more general equation (\ref{polariz:eq7bSedov}), the ratio $q_2P\rs{w}^2/q_0$ increases downstream with distance from the shock. Therefore, the condition $q_2P\rs{w}/q_0\ll 1$ could not be valid at radii lower than some distance from the center. Nevertheless, the differences are not large and the approximate solution may be used if one needs to reduce the computational cost. 

At a first glance, it seems that a similar approximate solution might be obtained for the limiting case B. 
In fact, this case assumes equal rates at the shock. If the growth and damping would be almost the same everywhere downstream then the equation would be much simpler: $d\bar P\rs{w}/d\bar a+\bar q_1\bar P\rs{w}=0$. However, the ratio of the damping to growth rates ($\alpha^{-1}$) increases downstream quite quickly with decrease of $\bar a$ (less waves survived) preventing us from such a simplification. 

Calculations show that profiles $\bar P\rs{w}(\bar r)$ are almost insensitive to the value of $q$. 
Smaller values of $b$ results in higher $\bar P\rs{w}$ in the SNR interior because the saturated level $P\rs{wsi}\propto \bar a^{3b/2}$ ($\bar a\leq 1$) was larger at previous times for smaller values of the parameter $b$. 
In order to reduce the parameter space, $b$ and $q$ are set to zero in our calculations.

%--------------------------------------------------------------
%--------------------------------------------------------------

%--------------------------------------------------------------
\begin{figure*}
 \centering
 \includegraphics[width=14.8truecm]{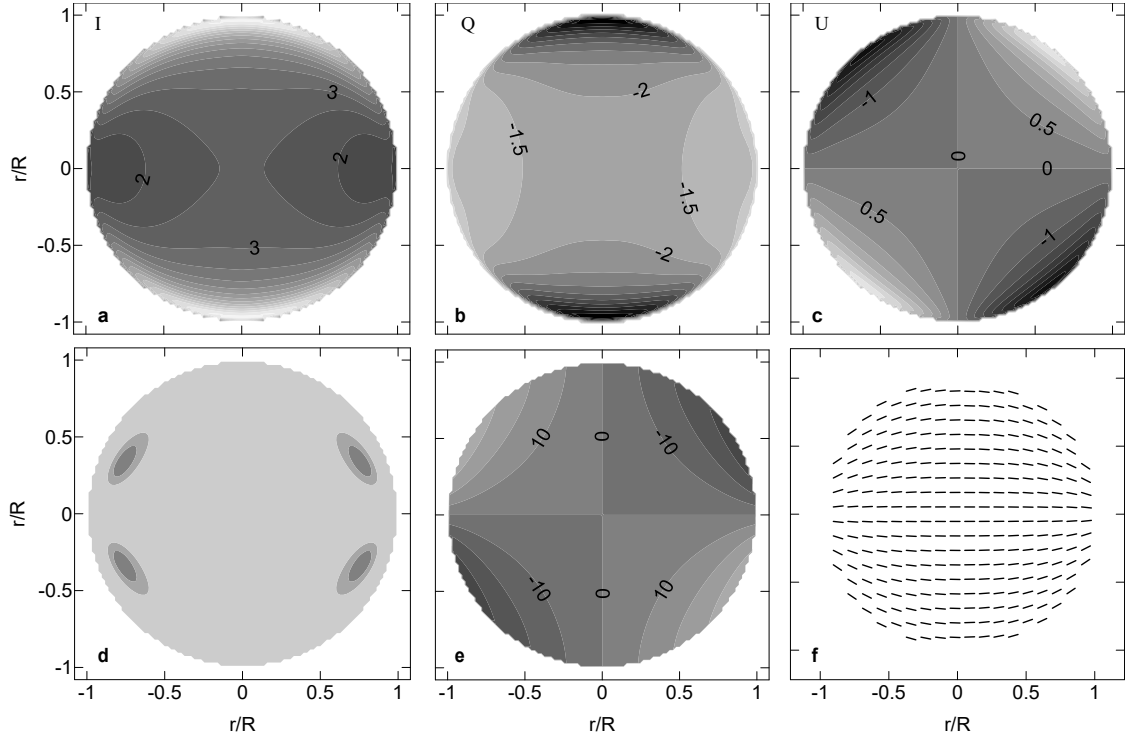}
 %\vspace{5truecm}
 \caption{Maps of the Stokes parameter $I$ (surface brightness) ({\bf a}), $Q$ ({\bf b}), $U$ ({\bf c}), polarization fraction $\Pi$ ({\bf d}), magnetic polarization angle $\Psi$ ({\bf e}) and vector map of `MF orientations' ({\bf f}) for a Sedov SNR with completely ordered MF (projected along the horizontal axis) and no Faraday rotation. The aspect angle is $\phi\rs{o}=90^\mathrm{o}$ (i.e. $\mathbf{B}\rs{o}$ is in the projection plane). Grayscale limits (the smaller the darker): (a) from $0.5$ to $9.5$, step $0.5$ in arbitrary units, (b) from $-6.5$ to $0.5$, step $0.5$ in arbitrary units, (c) from $-4.5$ to $4.5$, step $0.5$ in arbitrary units, (d) from $0.6895$ to $0.6905$, step $0.0005$, (e) min $-90$, max $90$, step $10$, in degrees. Hereafter, the length of lines in the MF direction maps are proportional to $\Pi$ with the maximum length for $0.7$. In this figure, $\langle\Pi\rangle=0.692$. 
               }
 \label{polariz:Bmapsa}
\end{figure*}
%--------------------------------------------------------------
%--------------------------------------------------------------
\begin{figure*}
 \centering
 \includegraphics[width=14.8truecm]{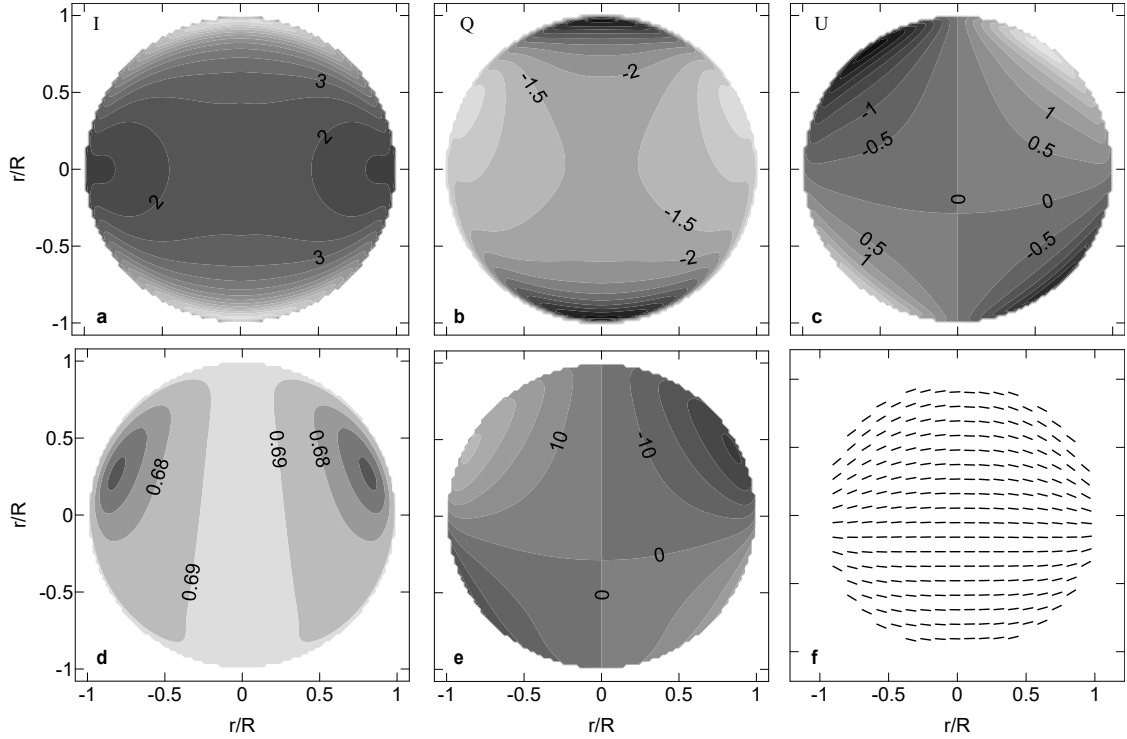}
 %\vspace{5truecm}
 \caption{The same as on Fig.~\ref{polariz:Bmapsa} but including the internal Faraday rotation, $\beta\rs{n}=1$. Grayscale for (d): min $0.64$, max $0.7$, step $0.01$. $\langle\Pi\rangle=0.686$.
               }
 \label{polariz:Bmapsb}
\end{figure*}
%--------------------------------------------------------------
%--------------------------------------------------------------
\begin{figure}
 \centering
 \includegraphics[width=6.4truecm]{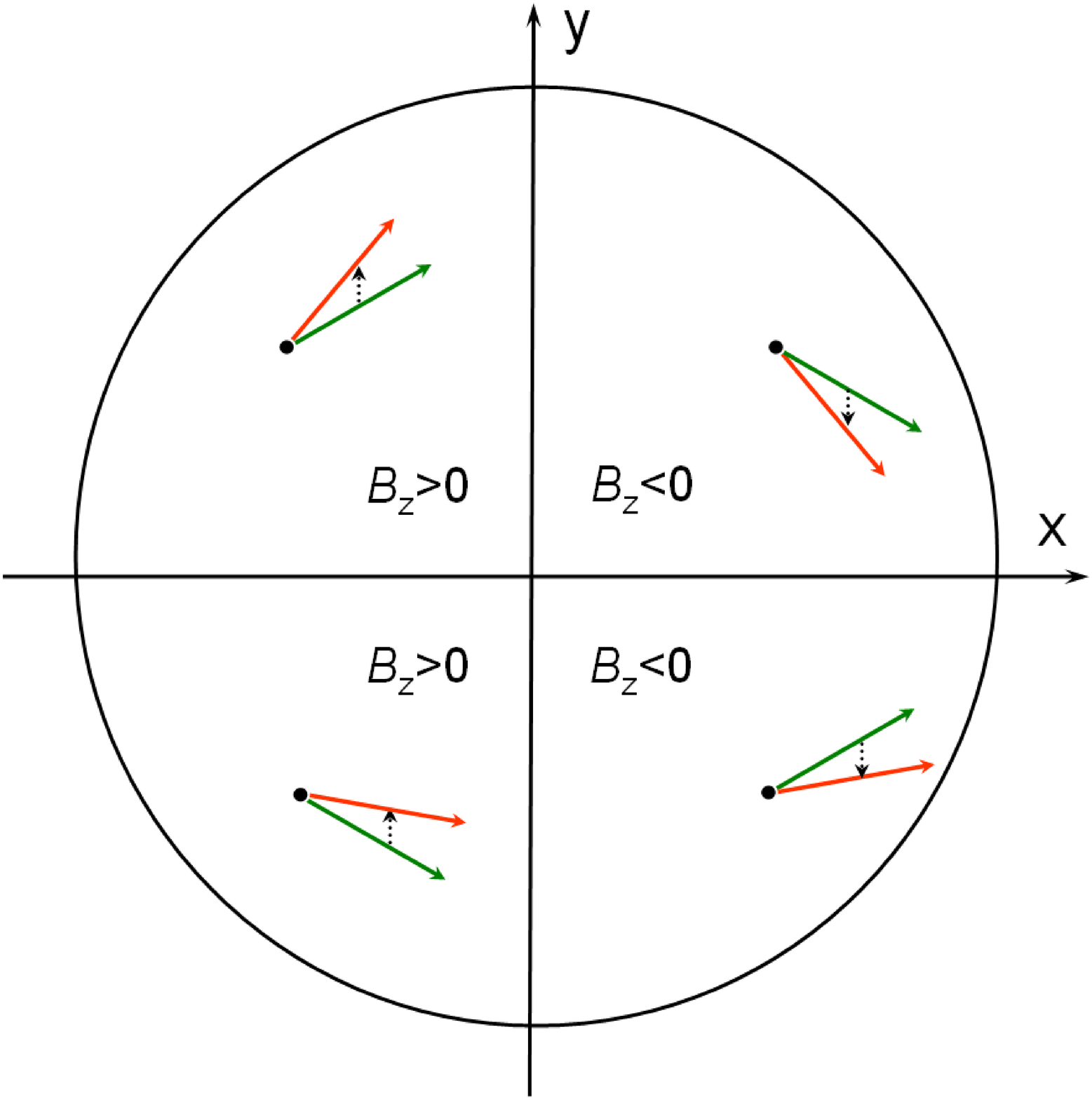}
 %\vspace{5truecm}
 \caption{{The schematic representation of the MF orientations, from the observer point of view: without (green) and with (red) the internal Faraday rotation; cf. Fig.~\ref{polariz:fig_pic}}.
               }
 \label{polariz:fig_pic_2}
\end{figure}
%--------------------------------------------------------------

%%%%%%%%%%%%%%%%%%%%%%%%%%%%%%%%%%%%%%%%%%%%%%%%%%%%%%%%%%%%%%%%%%%%%%%%%
\section{Polarization maps for Sedov SNRs}
\label{polariz:sect-results}

In this section we present the results of numerical simulations for SNRs evolving in the Sedov regime (adiabatic expansion in the medium with the uniform distribution of density and magnetic field). For these simulations we have used the assumptions and formulae as described in the above sections. In all our images, the projected ambient MF is along the horizontal $x$ axis. 
{Thanks to the semi-analytic nature of our simulations (Sedov solutions are analytic) we are not confined by the fixed grid of the numerical simulations with limited resolution in three dimensions. We can calculate all the necessary parameters in any point inside SNR and we may have as large resolution as we need. In practice, all our images have $100\times 100$ pixels in the projection plane and very large resolution (determined automatically by the integration algorithm) along the line of sight.}

\subsection{Completely ordered magnetic field}
\label{polariz:ordfieldresults}

For the sake of comparison with what will follow, let us present first the results of numerical calculations for the case of a fully ordered MF.

Fig.~\ref{polariz:Bmapsa} shows maps of the Stokes parameter $I$ (surface brightness), $Q$, $U$, the polarization fraction $\Pi$, the magnetic polarization angle (angle of `projected' MF with the $x$ axis) $\Psi$ and the associated vector map of the `MF orientations' inferred from the projected Stokes parameters, in the absence of the internal Faraday rotation ($\beta=0$), for the aspect angle (between the ambient MF direction and the line of sight) $\phi\rs{o}=90^\mathrm{o}$. The most prominent features are the quite symmetric patterns for $Q$ and $U$, the smoothly ordered `projected' MF and the high level of polarization: it is close to the maximum theoretical value, $\simeq 0.69$, over the whole projection. 

Fig.~10 of Paper I shows the same set of images but simulated in the {\it thin-layer approximation}. The maps from the present {\it three-dimensional model} (Fig.~\ref{polariz:Bmapsa}) resemble those images {rather} well. Some minor differences {can be noticed,} especially closer to the SNR edge: {they are essentially consequence of the fact that we now model in details the}  internal structure of SNR {which was neglected in the thin-shell approximation  adopted in Paper I}. In particular, {the most apparent differences are the} four 'islands` on the map of the polarization fraction, {which were not present in the same image in Paper I. They are real, not just a numerical effect, and the reason is the following: in} the case of the thin layer approximation, all the material is in an infinitely-thin layer around the SNR shock and there are only two emitting 'points` along each line of sight with different orientations of the MF vectors {(in the case of an aspect angle of $90^\mathrm{o}$ the orientations of the MF vectors are aligned on the two points)}. In contrast, the present model deals with different MF orientations in each point in the SNR interior. The depolarization on the $\Pi$ image is a bit higher where MF vectors vary to larger extent along the line of sight. In other words, {it is an effect of the MF jiggling along the line of sight, and} $\Pi$ thus is a 'measure' of the MF disorder along {this line}.

Fig.~\ref{polariz:Bmapsb} shows the same maps, calculated including the internal Faraday rotation. The polarization fraction is still high on Fig.~\ref{polariz:Bmapsb} but the lowest value is now $\Pi=0.522$ that is smaller than on Fig.~\ref{polariz:Bmapsa}. However, the polarization fraction is still high over the most of the SNR projection, as it is demonstrated by the average value of $\Pi$, $\langle\Pi\rangle=0.686$.  

Also in this case, {the} patterns {shown in} Paper I (see Fig.~14 there) {are reasonably reproduced. The most apparent} differences are again in the $\Pi$ map. {The regions with the lowest polarization fraction are not lying on the projected boundary, but slightly inside. In addition, this map is not symmetric any more with respect to the horizontal axis: also in this case, because (differently from the thin layer case) the level of the MF jiggling along the line of sight is now no longer symmetric.}
The internal Faraday effect causes the asymmetry between the upper and lower half of the $Q$ and $U$ maps. This translates in an asymmetry in the map of the polarization fraction. In fact, this effect is not present in the corresponding figures in Paper I {(Fig.~ 14 there)} which were simulated in the limit of the infinitely thin layer. 

{The pattern of this asymmetry depends on the orientation of the ambient MF. In our simulations, it is directed towards the right. Therefore, in the nearer half-sphere of SNR (i.e. for $z\geq 1$, see Fig.~\ref{polariz:fig_pic}), the line-of-sight component of the ordered MF ($B\rs{z}$) is preferentially positive (toward the observer) in the left half and preferentially negative in the right half of image. The orientation of $B\rs{z}$ in the nearer half-sphere of SNR is dominant for the internal Faraday rotation (the emission from the rearer part also passes through this SNR part). Therefore, the rotation is preferentially counterclockwise to the left (from the observer point of view) and clockwise to the right (Fig.~\ref{polariz:fig_pic_2}). The polarization planes are rotated by the same angle in the points symmetric with respect to the horizontal axis. However, the plane-of-the-sky components of MF have different orientation in the upper and lower half of image. As a result, the upper-lower asymmetry appears.}

{The degree of the asymmetry depends on the 'strength' of the internal Faraday effect.}
Eqs.~(\ref{polariz:betadef}) and (\ref{polariz:RMdef}) show that, due to the scaling property of the Sedov solutions, the angle of internal rotation $\beta$ in each point of the projection is proportional to the same product
\begin{equation}
 \beta\rs{n}=n\rs{o}B\rs{o\mu}R\rs{pc}\lambda\rs{m}^2
 \label{polariz:betandefSedov}
\end{equation}
where, $B\rs{o\mu}$ is the strength of the ambient MF in $\un{\mu G}$, $R\rs{pc}$ is the radius of SNR in pc, $\lambda\rs{m}$ is the wavelength in m. The parameter is, for example, $\beta\rs{n}=2.7$ for $n\rs{o}=1\un{cm^{-3}}$, $B\rs{o}=3\un{\mu G}$, $R=10\un{pc}$, $\lambda=30\un{cm}$. 
Thus, the parameter $\beta\rs{n}$ determines how strong is the Faraday effect in the SNR interior.

%--------------------------------------------------------------
\begin{figure*}
 \centering
 \includegraphics[width=15truecm]{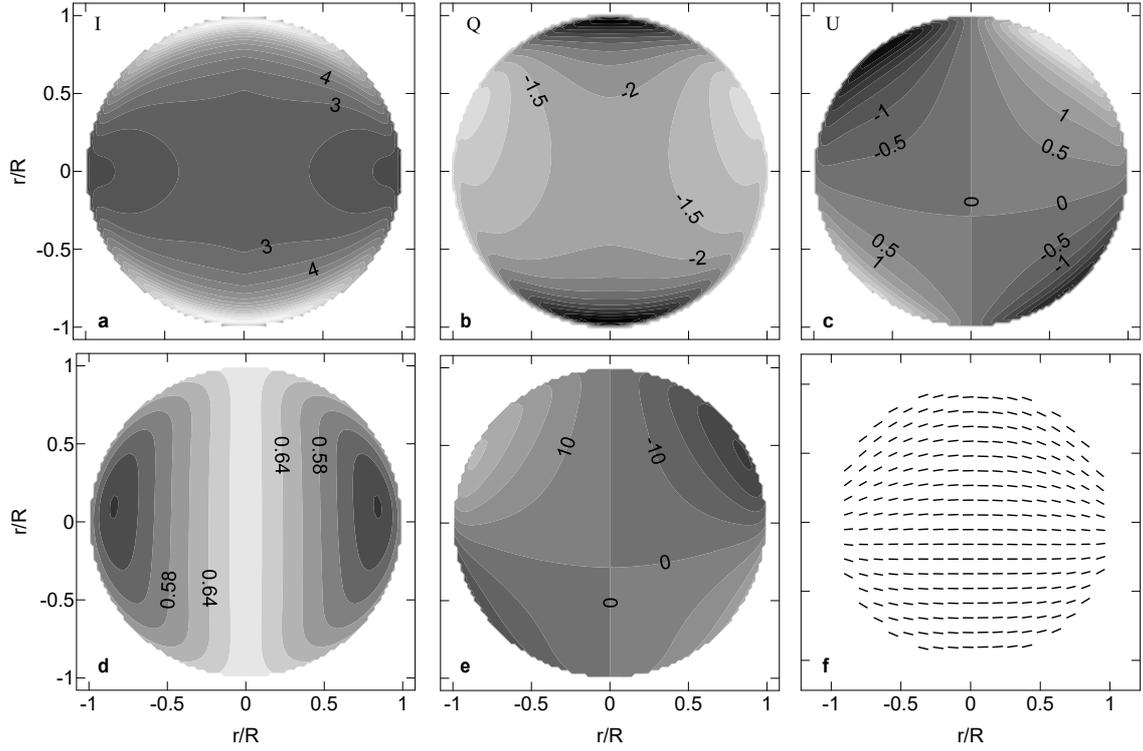}
 %\vspace{5truecm}
 \caption{The same as on Fig.~\ref{polariz:Bmapsa} but including the internal Faraday rotation and the disordered MF. Model A from Table~\ref{polariz:table1}. Grayscale for (d): min $0.46$, max $0.7$, step $0.03$. $\Pi\rs{min}=0.488$, $\langle\Pi\rangle=0.602$.
               }
 \label{polariz:Bmapsc}
\end{figure*}
%--------------------------------------------------------------

\subsection{Models with the partially disordered magnetic field}
\label{polariz:disordfieldresults}

In this subsection, we present maps of polarized emission in models which consider both the ordered $B$ and disordered $\delta B$ MF components. The  parameters for a number of considered models are summarized in Table~\ref{polariz:table1}. The basic set of parameters is that for the model labelled as A. Other models have one, or at most two, parameters changed with respect to model A and only these parameters are displayed in the table. 

%--------------------------------------------------------------
\begin{table}
 \centering
 \caption{Sets of parameters for the numerical simulations. %The values of $b$ and $q$ are zero for each set. 
The parameter $\beta\rs{n}$ is defined by Eq.~(\ref{polariz:betandefSedov}).
Set A is the reference model; while the other models differ from the case A only for the values shown in the table. `iso' means isotropic injection (i.e. independent of the shock obliquity), while `qpar' means quasiparallel injection (i.e. mostly at the parallel shocks).
         }
 \begin{tabular}{ccccc}
	\hline\hline
   Model&$\beta\rs{n}$&injection&$\Omega\rs{o\|}$&$\displaystyle\left(\frac{\delta B}{B}\right)\rs{s\|}$\\
	\hline\hline
   A&1&iso&0.4&0.3\\
	\hline
   B&0.2&&&\\
	\hline
   C&0.5&&&\\
	\hline
   D&2&&&\\
	\hline
   E&5&&&\\
	\hline
   F&&qpar&&\\
	%\hline
  % G&&&4&\\
	\hline
   G&&&10&\\
	\hline
   H&&&&1\\
	\hline
   I&&&10&1\\
	\hline\hline
 \end{tabular}
 \label{polariz:table1}
\end{table}
%--------------------------------------------------------------

Fig.~\ref{polariz:Bmapsc} shows the same set of images as on Fig.~\ref{polariz:Bmapsa} and Fig.~\ref{polariz:Bmapsb} but now including the Faraday rotation and the disordered magnetic field. We can see that the addition of the disordered MF (with $(\delta B/B)\rs{s\|}$ less than unity) does not alter considerably the images of the polarized emission. 

However, there are some differences visible. In particular, the surface brightness is somehow higher in the model which includes the partially turbulent MF: the area of the region with brightness between 2.5 and 3 units is larger on Fig.~\ref{polariz:Bmapsc}a comparing to Fig.~\ref{polariz:Bmapsb}a. This is because the Stokes parameter $I$ is higher if consider the turbulent MF, Eq.~(\ref{eq:IPL}); this effect should be more prominent if $\delta B/B$ is larger than unity (Fig.~\ref{polariz:fig_pol}). Another feature is visible on the same maps, in the shape of the grayscale border marked by the value 3: on Fig.~\ref{polariz:Bmapsc}a it is like $\}$ while it is like $)$ on Fig.~\ref{polariz:Bmapsb}a. The reason of this difference is that the ratio $\delta B/B$ is smaller in the perpendicular shock compared to the parallel one: the shock compression factor for $B$ increases from 1 to 4 respectively while $\delta B$ depends on the cosine of the obliquity angle, Eq.~\ref{polariz:Pwsfin-caseA}; thus $\delta B\rightarrow 0$ toward the perpendicular shock (which is projected as a vertical diameter) and the brightness decreases to the value it has without the turbulent MF.  

%--------------------------------------------------------------
\begin{figure*}
 \centering
 \includegraphics[width=17.4truecm]{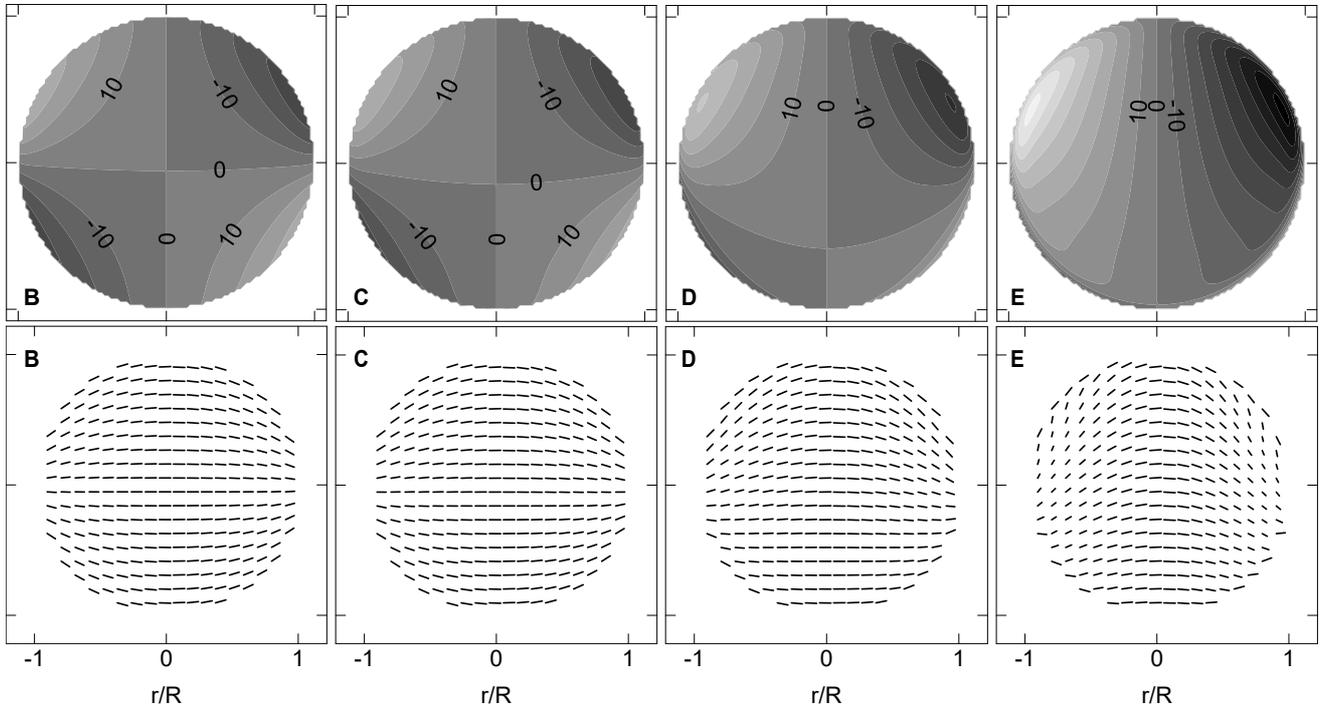}
 \caption{Maps of $\Psi$ (upper panels) and of the inferred `directions of MF' (lower panels) for different $\beta\rs{n}$ (models B-E, marked on the respective plots; to be compared to the model A, Fig.~\ref{polariz:Bmapsc}e,f). Aspect angle $\phi\rs{o}=90^\mathrm{o}$. $\Pi\rs{min}$ and $\langle\Pi\rangle$ are respectively: 0.505 and 0.607 (B), 0.502 and 0.606 (C), 0.439 and 0.588 (D), 0.183 and 0.501 (E). 
               }
 \label{polariz:Bmapsd}
\end{figure*}
%--------------------------------------------------------------
%--------------------------------------------------------------
\begin{figure*}
 \centering
 \includegraphics[width=17.4truecm]{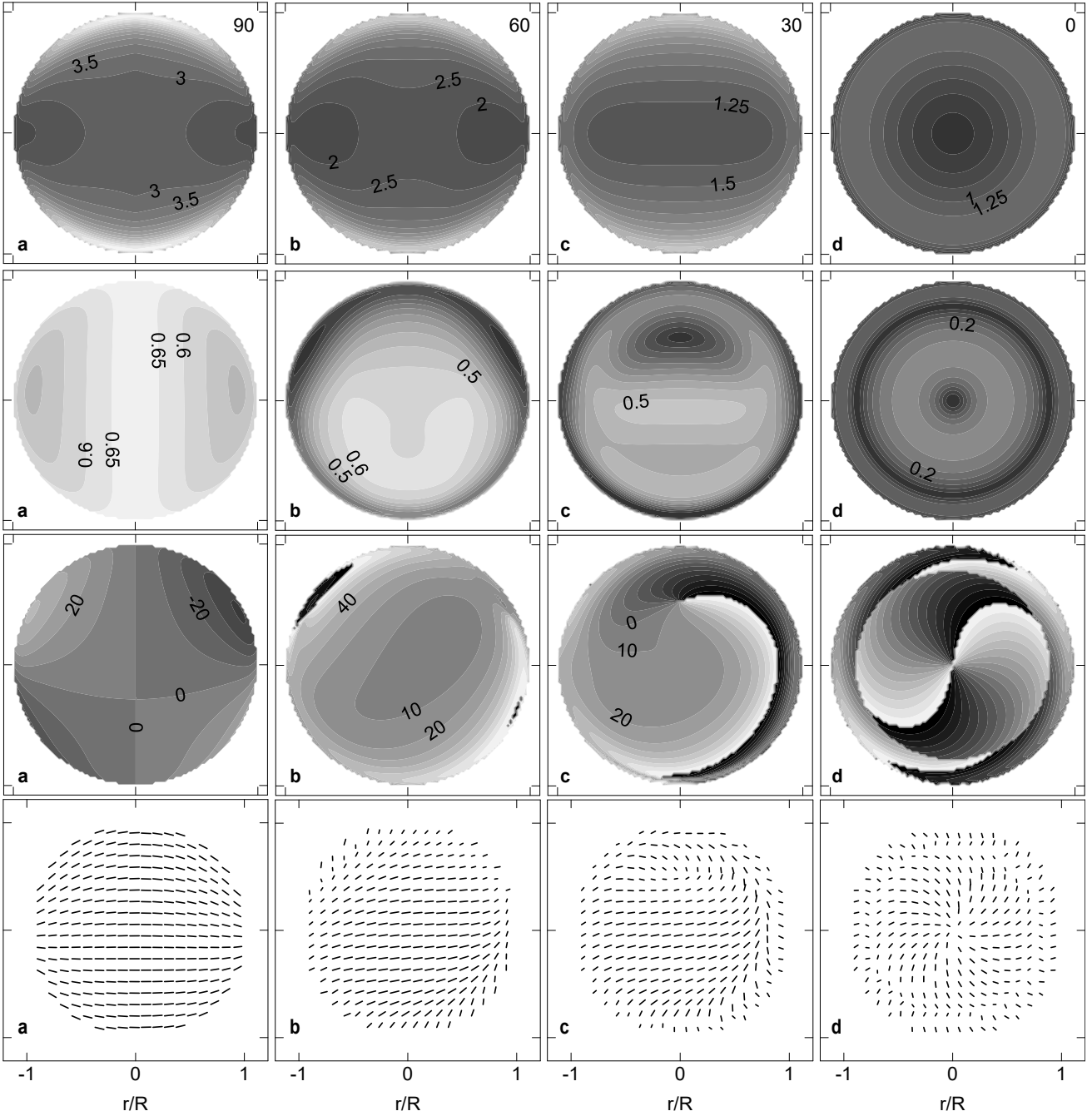}
 \caption{Maps of $I$ (upper panels), $\Pi$, $\Psi$ and inferred `directions of MF' (lower panels) for model A and different aspect angles $\phi\rs{o}$: {\bf a} -- $90^\mathrm{o}$, {\bf b} -- $60^\mathrm{o}$, {\bf c} -- $30^\mathrm{o}$, {\bf d} -- $0^\mathrm{o}$. Component of ISMF in the projection plane is along the horizontal axis. 
Scale for $I$: min 0.5 (black), max 9.5 (white), step 0.5 (a and b), min 0.25, max 4.75, step 0.25 (c and d), in arbitrary units.
Scale for $\Pi$: min 0 (black), max 0.7 (white), step 0.05. 
Scale for $\Psi$: min $-90^\mathrm{o}$ (black), max $90^\mathrm{o}$ (white), step $10^\mathrm{o}$. 
$\Pi\rs{min}$, $\langle\Pi\rangle$ and $\Pi\rs{max}$ are respectively: 0.488, 0.602, 0.691 (a), 0.007, 0.428, 0.642 (b), 0.001, 0.331, 0.579 (c), 0.000, 0.187, 0.317 (d). 
               }
 \label{polariz:Bmapse}
\end{figure*}
%--------------------------------------------------------------
%--------------------------------------------------------------
\begin{figure*}
 \centering
 \includegraphics[width=17.4truecm]{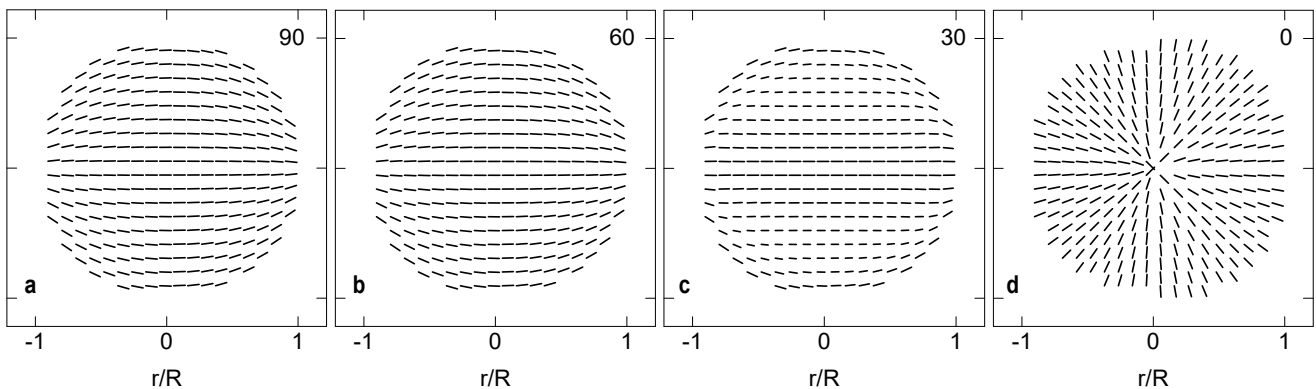}
 \caption{The same as lower panels at Fig.~\ref{polariz:Bmapse}, for $\beta\rs{n}=0$. 
               }
 \label{polariz:Bmapse0}
\end{figure*}
%--------------------------------------------------------------
%--------------------------------------------------------------
\begin{figure*}
 \centering
 \includegraphics[width=17.4truecm]{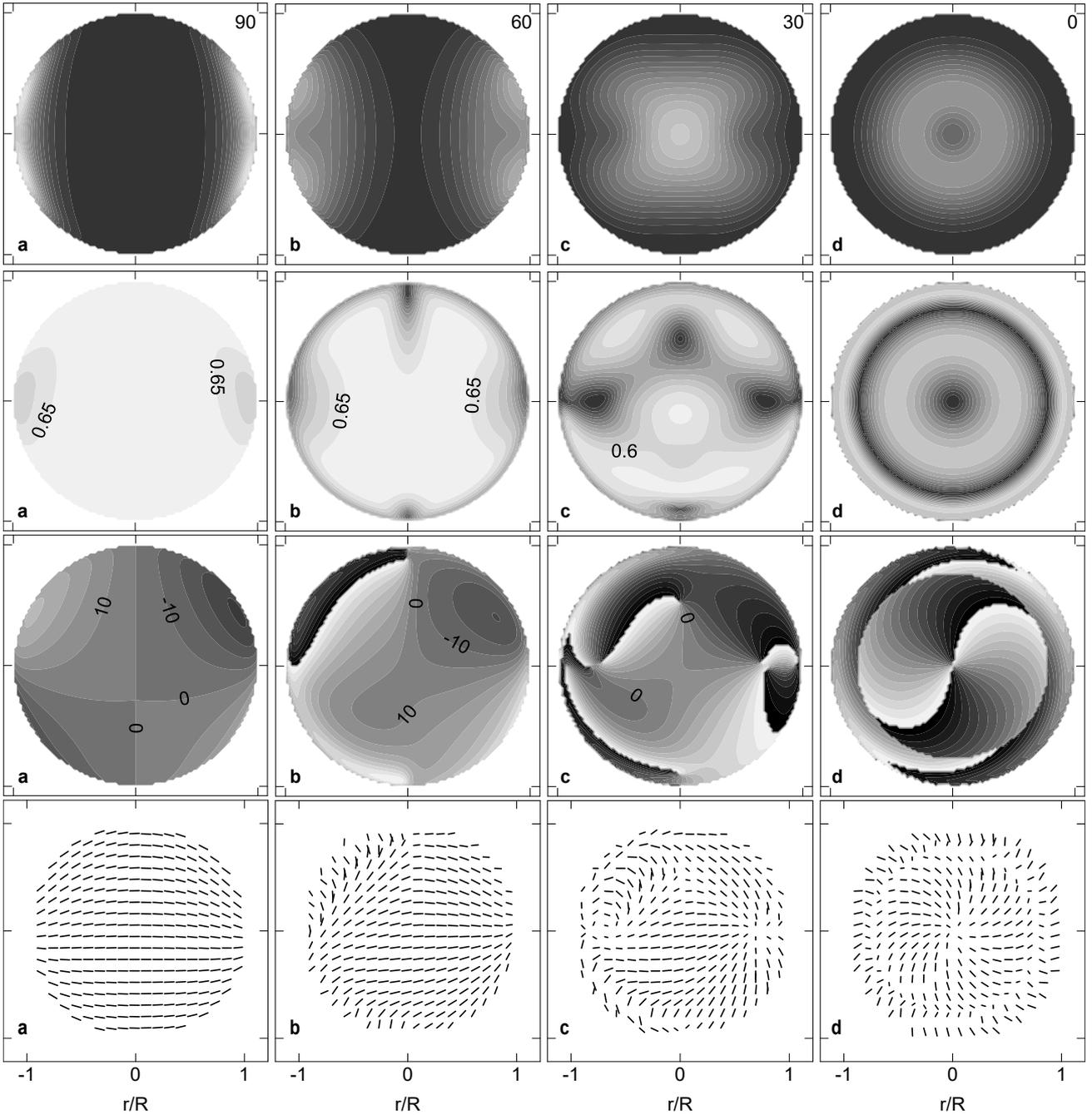}
 \caption{The same as Fig.~\ref{polariz:Bmapse}, for the quasiparallel injection (model F).
 Min, max and step for $I$ are 5 times (a) and 10 times (b-d) less than on Fig.~\ref{polariz:Bmapse}.
 Grayscale for $\Pi$ is min 0, max 0.7, step 0.05.
 $\Pi\rs{min}$, $\langle\Pi\rangle$ and $\Pi\rs{max}$ are respectively: 0.565, 0.675, 0.691 (a), 0.062, 0.605, 0.691 (b), 0.005, 0.474, 0.665 (c), 0.000, 0.391, 0.570 (d). 
               }
 \label{polariz:Bmapsf}
\end{figure*}
%--------------------------------------------------------------
%--------------------------------------------------------------
\begin{figure*}
 \centering
 \includegraphics[width=17.4truecm]{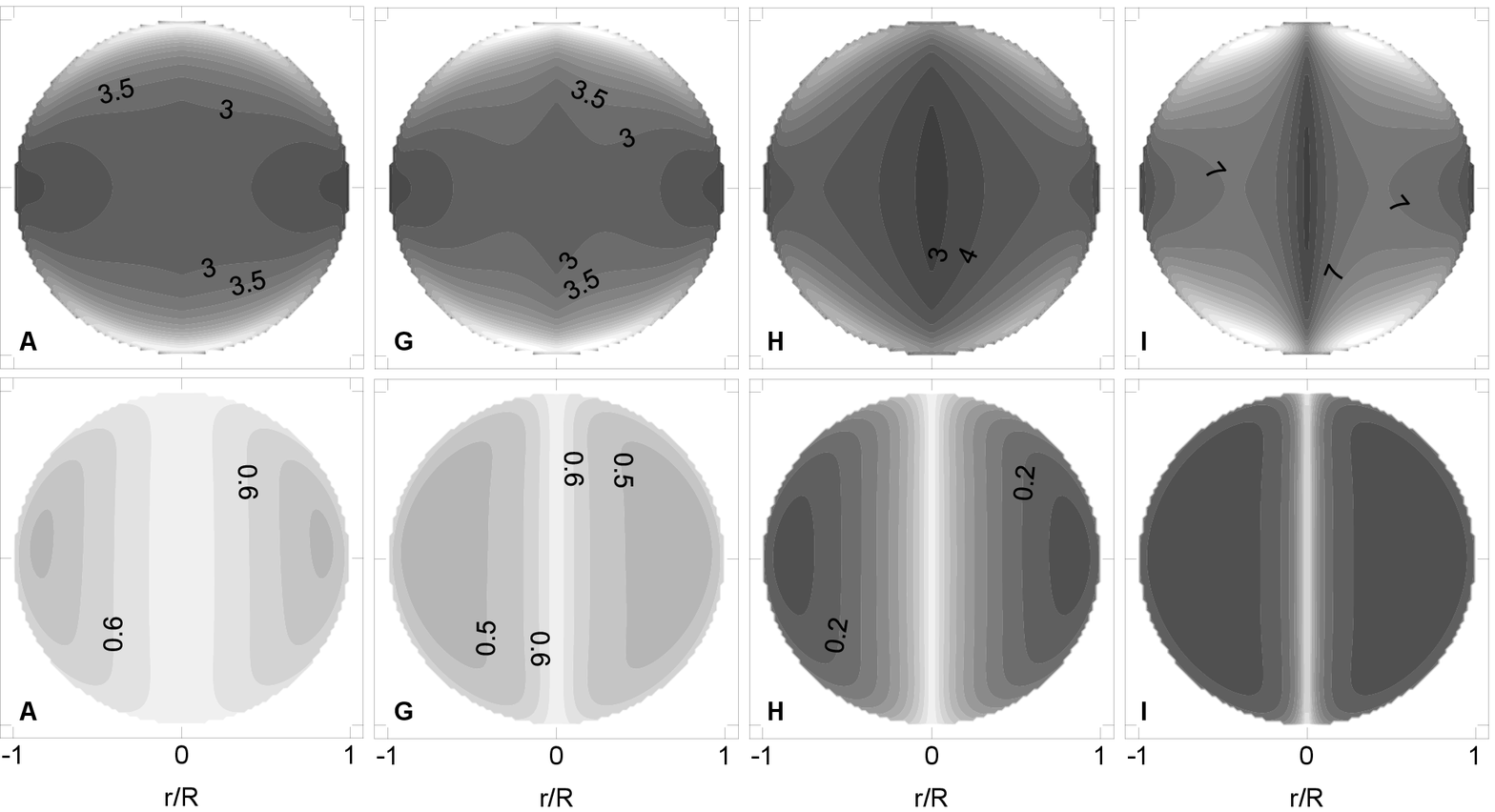}
 \caption{Maps of $I$ (upper panels) and $\Pi$ (lower panels) for models A, G, H, I (marked on the plots). $\phi\rs{o}=90^\mathrm{o}$.
 The only maps of $I$ and $\Pi$ are presented because other images ($Q$, $U$, $\Psi$) are very similar in all these models. 
 Min, max and step for $I$ in {\bf a} and {\bf b} are the same as on Fig.~\ref{polariz:Bmapsa}; they are 2 times larger in {\bf c} and {\bf d}.
 Grayscale for $\Pi$: min $0$, max $0.7$, step $0.05$. 
 $\Pi\rs{min}$, $\langle\Pi\rangle$, $\Pi\rs{max}$ are: 0.488, 0.602, 0.691 (A), 0.456, 0.533, 0.686 (G), 0.127, 0.309, 0.678 (H), 0.105, 0.181, 0.626 (I). 
               }
 \label{polariz:Bmapsg}
\end{figure*}
%--------------------------------------------------------------

More prominent differences are in the map of the polarization fraction (Fig.~\ref{polariz:Bmapsc}d, to be compared with Fig.~\ref{polariz:Bmapsb}d). As expected, the disordered MF depolarizes emission: the area of highly polarized regions is considerably smaller; these regions are located where the perpendicular shock is projected with $\delta B\rightarrow 0$. The values of $\Pi\rs{min}=0.488$ and $\langle\Pi\rangle=0.602$ shows how strong is this effect (to be compared to the maximum possible   value $0.692$). The polarization fraction is smaller for the same model if the angle $\phi\rs{o}$ between ISMF and the line of sight is smaller. In particular, $\Pi\rs{min}=0.007$ and $\langle\Pi\rangle=0.428$ for $\phi\rs{o}=60^\mathrm{o}$. 

As we already noticed in Sect.~\ref{polariz:ordfieldresults}, the internal Faraday effect changes the measured polarization directions and the overall effect of FR on SNR images depends on a parameter $\beta\rs{n}$, Eq.~(\ref{polariz:betandefSedov}). The role of $\beta\rs{n}$ in determining the pattern of the `observed MF directions' is apparent from Fig.~\ref{polariz:Bmapsd} where these patterns are calculated for models B-E in Table~\ref{polariz:table1} (see also model A on Fig.~\ref{polariz:Bmapsc}e,f).\footnote{The image of $I$ has the same pattern (though with different amplitude) for the models A-E because $\beta\rs{n}$ does not affect the contrasts in the map of this Stokes parameter.} 
The polarization images are sensitive to $\beta\rs{n}\geq 0.5$. 
%With increase of $\beta$ over $\approx 25$ the pattern of polarization directions might look like almost chaotic. 
The larger the parameter $\beta\rs{n}$ the stronger the effect of the internal Faraday rotation. 

\citet{Dickel-Milne-1976} have suggested that a change of the `inferred MF' directions from the radial to the mostly tangential one could be the evolutional trend from young to old SNRs.  
Since $R$ is increasing with the SNR age, {we checked if} such temporal trend {is} visible with increase of $\beta\rs{n}$ {in our model. In fact}, there is no clear trend from preferentially radial {to tangential} polarization directions {in our model} (Fig.~\ref{polariz:Bmapsd}). 
{Since we consider here the Sedov phase only, our results agree with the expectation of \citet{Dickel-Milne-1976} that the change in magnetic-field orientation is `possibly corresponding to the first and third evolutionary phases of an SNR discussed by \citet{Woltjer-1970}.' In fact,} i) we do not consider ejecta (Sedov model is just the point explosion) and thus the MHD instabilities which might result in the radial `fingers' in young SNRs \citep[e.g.][]{Orlando-etal-2012} are not accounted for; ii) the radiative losses of the shocked plasma (efficient after the Sedov stage) increase the tangential component of MF in old SNRs \citep{petr-2016}. Future simulations have to be extended to the SNR ages before and after the adiabatic phase in order to reconsider the idea of \citet{Dickel-Milne-1976} which seems to be confirmed by recent observations \citep{Dubner-Giacani-2015}. 

Another observed property -- decrease of the polarization at lower frequencies -- is restored in our simulations. In fact, the polarization is smaller for larger $\beta\rs{n}$ (Fig.~\ref{polariz:Bmapsd}, look at values of $\Pi\rs{min}$ and $\langle\Pi\rangle$) and  $\beta\rs{n}$  inversely depends on the frequency $\beta\rs{n}\propto \lambda^2\propto \nu^{-2}$.  

Till now, we have considered the SNR projections for the aspect angle $\phi\rs{o}=90^\mathrm{o}$, i.e. when the uniform ISMF lies in the projection plane. The polarization is smaller for other orientations of the ambient MF versus the observer (Fig.~\ref{polariz:Bmapse}). We see again that the main factor determining the polarization patterns is the internal Faraday rotation. Our simulations demonstrate that the SNR images do not generally reflect the actual orientation of MF inside SNR because they gives the polarization directions of emission affected by the internal Faraday rotation. {\it The larger the role of the internal Faraday rotation (i.e. the higher $\beta\rs{n}$), the less similar is the observed polarization pattern to the actual orientation of MF in SNR.} Therefore, observed patterns may not be used to extract the MF orientations in SNR until it is proved that the internal Faraday rotation is negligible (i.e. that $\beta\rs{n}<0.5$; for example, observing at a high enough frequency, the Faraday rotation becomes asymptotically negligible, see e.g. the moments approach in Paper I). This is clearly seen by comparing lower panels on Fig.~\ref{polariz:Bmapse} and Fig.~\ref{polariz:Bmapse0}; in the latter one, the images are calculated for the value $\beta\rs{n}=0$ and the maps trace therefore the orientation of MF. 

It is visible on our images that the Faraday effect could be responsible for the spatial domains in the maps with abrupt change of the polarization directions. This is a feature observed in many SNRs \citep[e.g.][]{Dubner-Giacani-2015}. %The area of domains with essentially different directions of polarization planes vary with orientation of SNR and ISMF versus the observer.

If injection preferentially happens at the parallel shocks (model F) then the images of $I$ and of the polarized emission $\sqrt{Q^2+U^2}$ have maxima located around the parallel shock regions,  where emitting particles reside (Fig.~\ref{polariz:Bmapsf}). 
Some increase of the polarization fraction from the SNR as a whole is prominent comparing to the isotropic injection case (compare values of $\Pi\rs{min}$, $\langle\Pi\rangle$ and $\Pi\rs{max}$ given in the figure captions). 

How do the properties of the turbulent MF affect images of polarized emission of SNRs? Two parameters in our model regulate behavior of waves in SNR. The first one, $\Omega\rs{o\|}$, is $\Omega$ immediately before the parallel shock. This parameter, defined by Eq.~(\ref{polariz:defOmega}), determines the ratio between the rates of the wave growth and damping  (see Eq.~\ref{polariz:alphadef}). The value $\Omega\rs{o\|}=0.4$ in the basic model A brings us to the limiting case of the small wave damping (Fig.~\ref{polariz:fig_alphalim}). In this regime the pressure of waves decreases to zero toward the regions of the perpendicular shock, Eq.~(\ref{polariz:Pwsfin-caseA}). The role of the fluctuating field is therefore prominent mostly around the regions with the parallel shocks (i.e. around `poles' which are on the left and right edges of images with $\phi\rs{o}=90^\mathrm{o}$). Another limit of the wave evolution, when the growth and damping of waves have the same rates, is represented by the model G. Now, the ratio of the rates does not depend on the obliquity. Thus, the effect should lead to overall decrease of the polarization fraction (because of increase of the volume with turbulent MF) and morphological differences, mostly where the regions with the perpendicular shocks are projected. This is what we see on Fig.~\ref{polariz:Bmapsg} (models A and G). 

The second parameter which determines properties of the waves is the ratio $\left(\delta B/B\right)\rs{s\|}$. We have seen that the differences between images with and without fluctuating component of MF are rather small. The reason is that the role of the disordered MF on that SNR images is not strong because of the small value $\left(\delta B/B\right)\rs{s\|}=0.3$. Such value was taken because our model assumes that the nature of the fluctuating MF is Alfv\'en waves. Theory for these waves implicitly assumes $\delta B\ll B$. Nevertheless, in order to see what could be the effect of the stronger waves, we run simulations also for models H and I, with $\left(\delta B/B\right)\rs{s\|}=1$ (Fig.~\ref{polariz:Bmapsg}). These simulations are done just to imagine an  effect of efficient disordered MF on the polarized emission from SNRs. The main effects are the increase of the surface brightness with increase of $\delta B/B$ and the decrease of the polarization fraction, in agreement with Fig.~\ref{polariz:fig_pol}.

{
The model in the present paper is based on a principal condition of unmodified shocks. To this extent, our model is self-consistent: we consider test-particle acceleration, Alfv\'en waves, no MF amplification in the precursor due to CR streaming. This is in agreement with our assumption of adiabatic SNRs which is typically applied to evolved SNRs where the CR back-reaction is not effective, as commonly agreed, and as we have shown in \citep{band-petr-2010} with our statistical analysis. We hope that in the future the effects of the nonlinear particle acceleration as well as other modes (including Bell's one) of waves, on the polarized images of SNRs may be considered, in a self-consistent way. We would like to note that there are almost no studies (see review in the Introduction) where the overall maps of the SNR surface brightness are considered with the non-linear effects included (though there are papers dealing with such effects locally, e.g. close to the shock only). In order to study these effects in the polarization images, one has to invest much more effort than to simulate the brightness maps.\footnote{{Though some aspects of the problem are studied, to our knowledge, there is neither model nor numerical simulations which consider both MF amplification and its evolution deep downstream in the whole SNR volume, for any obliquity. The reason is the different time and length scales: the amplification involves micro-physics which develops on the scale of the shock precursor while MF evolves downstream on the scales of the SNR radius. In the present paper, we consider the Alfv\'en waves because their theory is well developed.}} 
However, there is an approximate approach to consider some aspects of how the CR shock modification affects the polarization images. It is suggested by \citet{Ellison-etal-2004} and \citet{Ferrand-etal-2010} and is implemented, in particular, in our papers on SNR images \citep{xmaps,Orlando-etal-2011}. It consists in the use of an 'effective' adiabatic index $\gamma=4/3$ or even $\gamma=1.1$ which  is intended to mimic respectively a moderate and a strong shock modification. With these indices the shock compression is higher: $\sigma=7$ and $\sigma=21$ respectively  (to be compared to the classical value $\sigma=4$ for $\gamma=5/3$). The consequences of such an approach are the radially-thinner features in the brightness images (e.g. Fig. 7 in \citet{Orlando-etal-2011} and approximate analytical formulation in \citet{xmaps}). In addition, the increased compression ratio decreases the electron spectral index $s$ ($s=1.5$ and $s=1.15$ respectively). This results in a small change of the Stokes parameters $Q$ and $U$ (Fig.1 of the Paper I) and $\Pi\rs{max}$ (according to Eq.~\ref{polariz:Pimaxdef}; it is $0.652$ and $0.617$ respectively while $\Pi=0.692$ for $\gamma=5/3$), and is completely ineffective to change the ratio $\Pi/\Pi\rs{max}$ (Fig.2 of the Paper I). Therefore, we expect that the non-linear acceleration will create higher-order corrections to our test-particle results which are able to catch the most prominent effects. }

%%%%%%%%%%%%%%%%%%%%%%%%%%%%%%%%%%%%%%%%%%%%%%%%%%%%%%%%%%%%%%%%%%%%%%%%%
\section{Conclusions}

The present paper is devoted to development of the model and to simulations of images of the polarized synchrotron emission from Sedov SNRs. The main goal is a general analysis of the polarization maps. Namely, we are interested to see how the properties of SNR, its internal structure and the turbulent component of MF affect the polarization patterns of these objects. We leave discussion on certain objects to future studies. 

Some aspects of the problem were considered by \citet{Schneiter-etal-2015}, in limits of the negligible internal Faraday rotation and of the zero turbulent component of MF. The approach we developed in Paper I and in the present paper accounts for both of these components. 

In order to model polarized emission from relativistic electrons in SNRs, {we have solved} the \textit{two essential problems}. First, the classical theory of synchrotron emission has to be generalized to describe the emissivity in the ordered plus disordered MF. A way to such generalization is presented in Paper I; it restores, in the limit $\delta B\ll B$, the known description. 
Second, the distribution of the disordered MF component has to be known inside the SNR interior. 
To this purpose, we developed a model for the evolution of the random component of MF downstream of the adiabatic shock under the assumption that it is due to the growth and damping of Alfv\'en waves. This type of waves is suitable because the whole task is complex in itself but the physics of the Alfv\'en waves is more or less known. 
In doing so, we consider the interactions of waves with particles accelerated by the SNR forward shock.
To this extend, our treatment is non-linear: evolution of waves are connected to relativistic particles. However, we do not consider possible influence of particles and waves on the MHD structure of SNR. The evolution of the random MF component is considered to be on the magneto-hydrodynamic `background' of Sedov shock. 

Having traced the 3-D distribution of all the relevant physical quantities inside SNR, we can consider the detailed spatial distribution of the polarized radio emission and synthesize the projected images of polarization parameters  under different SNR orientations with respect to the observer.

{Our assumption of adiabatic SNRs implicitly shifts the attention towards not very young SNRs, namely to cases in which the effects of the non-linear particle acceleration should not be important. Probably it could be worth specifying this: we miss some part of generality, but we are right in the frames of the clearly described model which may be applied to many SNRs.}

We have demonstrated that the random MF component reduces the polarization fraction, as expected. {A new effect is that} the presence of the random MF component increases the synchrotron emissivity (Fig.~\ref{polariz:fig_pol}). This produces important consequences for fitting the broadband SNR spectra. If, for example, $\delta B\sim B$ then the synchrotron flux is twice the flux for the case where the random MF is neglected. 

The role of the Faraday effect \textit{in the SNR interior} (i.e. the different rotation of the polarization plane from each point along the line of sight inside SNR) is important in formation of the Stokes parameter maps. The uniform derotation of the observed polarization maps might give a `true' image of MF only if the internal Faraday rotation is negligible, that is the exception rather than the rule. The uniform derotation corrects for the Faraday effect in ISM (between the SNR and the observer) but not for the internal rotation which vary essentially over the SNR projection. The pattern of the polarization maps is affected by the internal Faraday rotation if {the parameter} $\beta\rs{n}$ {(Eq.~\ref{polariz:betandefSedov})} is greater than $\simeq 0.5$.

Our simulations restore the observed properties: 
i) there are polarization domains with abrupt change of polarization direction in SNR maps; 
ii) the polarization fraction is small; 
iii) the polarization fraction decreases for larger $\lambda$. 

{As discussed in Sect.~\ref{polariz:ordfieldresults}, the simulated maps presented here show a general agreement with those already presented in Paper I. 
The main differences can be noticed close to the SNR projected edge, as a consequence of the fact that we have released here the thin-shell approximation. This leads essentially to two kinds of effects: i) the presence of a shell with a finite geometrical thickness leads to a shift of some features away from the edge, giving rise to features like, for instance, some `islands' now appearing in maps of the polarization fraction; ii) the fact that the direction of projected MF now can change along the line of sight, enhances the level of depolarization, and this effect is even more apparent in the cases with a low local random MF component.}

{In Paper I we have drawn the basis to treat the synchrotron emission in a partially ordered magnetic field, as well as to model the basic and most important geometrical aspects in a shell-type SNR. In the present paper we have added a treatment of the physics of particle acceleration and evolution, in a test-particle regime. The similarity between the results in these two papers should be seen as a proof of the validity of the simplified approach chosen for Paper I. The simpler thin-layer approach may be used in cases when one need an idea about the overall pattern of polarization images. When, however, the goal is to understand the observed patterns or to study some details in images then 3D simulations are needed. We believe that an advantage of our approach is that it could be implemented, with a moderate effort, to some real cases, with more complicate geometry.}

There are observational evidences \citep{Dickel-Milne-1976,Dubner-Giacani-2015} that patterns of polarization maps of the young SNRs demonstrate predominantly radial orientation of the projected MF directions while maps of the older SNRs have mostly tangential orientation. Our simulations suggest  that, in order to restore this effect numerically, the model has to go beyond the `point-explosion' approach and beyond the adiabatic stage, in order to treat the ejecta-ISM interactions in the young SNRs and the influence of the radiative losses of the shocked plasma on the magnetic field.

%%%%%%%%%%%%%%%%%%%%%%%%%%%%%%%%%%%%%%%%%%%%%%%%%%%%
\section*{Acknowledgements}

{We thank Pasquale Blasi and Elena Amato for interesting discussions and suggestions, as well as the referee for useful recommendations.}
This work is part of a project started with funds of PRIN INAF 2010. It is partially funded by the PRIN INAF 2014 grant `Filling the gap between supernova explosions and their remnants through magnetohydrodynamic modeling and high performance computing'. All simulations were performed on the computational cluster in Institute for Applied Problems in Mechanics and Mathematics, partially in the frame of the project 0117U002815.

%%%%%%%%%%%%%%%%%%%%%%%%%%%%%%%%%%%%%%%%%%%%%%%%%%%%%%%%%%%%%%%%%%%%%%%%%

%%%%%%%%%%%%%%%%%%%%%%%%%%%%%%%%%%%%%%%%%%%%%%%%%%%%%%%%%%%%

%%%%%%%%%%%%%%%%%%%%%%%%%%%%%%%%%%%%%%%%%%%%%%%%%%%%%%%%%%%%%%%%%%%%%%%%%%%%%%%%%%%%%%%%%%%%%%%%%
\appendix
\section[]{Components of magnetic field}
\label{polariz:appA}

In this Appendix, we present the formulae to calculate the components of magnetic field. 

Let us consider the reference frame with z-axis directed toward the observer. The ambient MF is in the yz plane (without loss of generality). The aspect angle $\phi\rs{o}$ is the angle between the ambient MF $\mathbf{B}\rs{o}$ and the line of sight. The corresponding spherical coordinates are $(r,\theta,\varphi)$. 

In each point, MF $\mathbf{B}$ is a vector sum of the radial $\mathbf{B}\rs{r}$ and the tangential $\mathbf{B}\rs{t}$ components which may be re-written as
$
 \mathbf{B}=\bar B\rs{r}\mathbf{B}\rs{ro}+\sigma \bar B\rs{t}\mathbf{B}\rs{to}
 =\bar B\rs{r}\mathbf{B}\rs{ro}+\sigma \bar B\rs{t}(\mathbf{B}\rs{o}-\mathbf{B}\rs{ro}),
$
where $\bar B\rs{r}=B\rs{r}/B\rs{rs}$, $\bar B\rs{t}=B\rs{t}/B\rs{ts}$, indices 's' and 'o' mark the immediate post- and pre-shock values, $\sigma$ is the shock compression factor. 
Therefore,
\begin{equation}
 \mathbf{B}=\sigma\bar B\rs{t}\mathbf{B}\rs{o}+(\bar B\rs{r}-\sigma\bar B\rs{t})(\mathbf{B}\rs{o}\cdot\mathbf{n})\mathbf{n}
\end{equation}
where $\mathbf{n}$ is the unit vector in the radial direction.
The projection of $\mathbf{B}$ on the Cartesian x axis is given by the dot product
\begin{equation}
 B\rs{x}=(\mathbf{B}\cdot\hat x)=\sigma\bar B\rs{t}{B}\rs{ox}+(\bar B\rs{r}-\sigma\bar B\rs{t}){B}\rs{o}\cos\Theta\rs{o}{n\rs{x}}
\end{equation}
where $\Theta\rs{o}$ is the shock obliquity angle. 
Expressions for $B\rs{y}$ and $B\rs{z}$ are the same but the index 'x' should be changed to 'y' and 'z' respectively. The Cartesian components of $\mathbf{n}$ are
$
 (\sin\theta\cos\varphi,\sin\theta\sin\varphi,\cos\theta).
$
The components of $\mathbf{B}\rs{o}$ are
$
 (0,B\rs{o}\sin\phi\rs{o},B\rs{o}\cos\phi\rs{o}).
$ 
Taking all these together, we have the MF components:
\begin{equation}
\begin{array}{l}
 B\rs{x}/B\rs{o}=(\bar B\rs{r}-\sigma\bar B\rs{t})\cos\Theta\rs{o}\sin\theta\cos\varphi,\\
 B\rs{y}/B\rs{o}=\sigma\bar B\rs{t}\sin\phi\rs{o}+(\bar B\rs{r}-\sigma\bar B\rs{t})\cos\Theta\rs{o}\sin\theta\sin\varphi,\\
 B\rs{z}/B\rs{o}=\sigma\bar B\rs{t}\cos\phi\rs{o}+(\bar B\rs{r}-\sigma\bar B\rs{t})\cos\Theta\rs{o}\cos\theta.
\end{array}
\end{equation} 
As to the obliquity angle, it is given by the dot product $\cos\Theta\rs{o}=(\mathbf{B}\rs{o}\cdot\mathbf{n})/B\rs{o}$ and is therefore  
\begin{equation}
 \cos\Theta\rs{o}=\sin\phi\rs{o}\sin\theta\sin\varphi+\cos\phi\rs{o}\cos\theta.\\
\end{equation}

%%%%%%%%%%%%%%%%%%%%%%%%%%%%%%%%%%%%%%%%%%%%%%%%%%%%%%%%%%%%%%%%%%%%%%%%%%%%%%%%%%%%%%%%%%%%%%%%%
\bsp
\label{lastpage}

\begin{thebibliography}{99}
\bibitem[{{Amato} \& {Blasi}(2005)}]{amato2005} Amato E., Blasi P., 2005, MNRAS 364, L76
\bibitem[{{Amato} \& {Blasi}(2006)}]{amato2006} Amato E., Blasi P., 2006, MNRAS 371, 1251
\bibitem[{{Amato} \& {Blasi}(2009)}]{amato2009} Amato E., Blasi P., 2009, MNRAS 392, 1591
%\bibitem[{{Ballet}(2006)}]{Ballet-2006} Ballet J. 2006, Adv. Space Res. 37, 1902
\bibitem[{{Balsara} \& {Kim}(2005)}]{Balsara-Kim-2005} Balsara D., Kim J. 2005, ApJ 634, 390
\bibitem[{{Bamba} {et al.}(2005)}]{Bamba-etal-2005} {Bamba A., Yamazaki R., Yoshida T., Terasawa T., Koyama K. 2005, ApJ 621, 793}
\bibitem[{{Bandiera} \& {Petruk}(2010)}]{band-petr-2010} Bandiera R., Petruk O. 2010, A\&A 509, A34
\bibitem[{{Bandiera} \& {Petruk}(2016)}]{band-petr-2016a} Bandiera R., Petruk O. 2016, MNRAS, 459, 178 (Paper I)
\bibitem[{{Bell}(2004)}]{Bell-2004} Bell A. 2004 MNRAS 353, 550
\bibitem[{{Berezhko} {et al.}(2003)}]{Ber-Volk-2003-mf} {Berezhko E., Ksenofontov, L., V\"olk H. 2003, A\&A 412, L11} 
\bibitem[{{Berezhko} \& {V\"olk}(2004)}]{Ber-Volk-2004-mf} {Berezhko E., V\"olk H. 2004a, A\&A 419, L27} 
\bibitem[{{Beshley} \& {Petruk}(2012)}]{ppmaps} Beshley V., Petruk O., 2012, MNRAS 419, 1421
\bibitem[{{Bocchino} {et al.}(2011)}]{Bocchino-et-al-2011} Bocchino F., Orlando S., Miceli M., Petruk O. 2011, A\&A 531, A129
\bibitem[{{Bykov} {et al.}(2008)}]{Bykov-et-al-2008} Bykov A., Uvarov Y., Ellison D. 2008, ApJ 689, L133
\bibitem[{{Bykov} {et al.}(2009)}]{Bykov-et-al-2009} {Bykov A., Uvarov Y., Bloemen J., den Herder J., Kaastra J. 2009, MNRAS 399, 1119}
\bibitem[{{Bykov} {et al.}(2011)}]{Bykov-et-al-2011} Bykov A., Ellison D., Osipov S., Pavlov G., Uvarov Y. 2011, ApJ 735, L40
%\bibitem[{{Caprioli} {et al.}(2009a)}]{caprioli2009a} Caprioli D., Blasi P., Amato E., Vietri M., 2009a, MNRAS 395, 895
\bibitem[{{Caprioli} {et al.}(2009b)}]{caprioli2009} Caprioli D., Blasi P., Amato E., 2009, MNRAS 396, 2065
\bibitem[{{Chevalier}(1974)}]{Chevalier1974} Chevalier R., 1974, ApJ, 188, 501
\bibitem[{{Cox} \& {Franko}(1981)}]{Cox-Franko-1981} Cox D.P., Franco J., 1981, ApJ 251, 687	
\bibitem[{{Dickel} \& {Milne}(1976)}]{Dickel-Milne-1976} Dickel J., Milne D., 1976, Aust.J.Phys. 29, 435
\bibitem[{{Dubner} \& {Giacani}(2015)}]{Dubner-Giacani-2015} Dubner G., Giacani E. 2015, Astron. Astroph. Rev. 23, 3
\bibitem[{{Ellison} {et al.}(2004)}]{Ellison-etal-2004} {Ellison D. et al. 2004, A\&A, 413, 189}
\bibitem[{{Eriksen} {et al.}(2011)}]{Eriksen-et-al-2011} Eriksen K., Hughes J., Badenes C. et al. 2011, ApJ 728, L28
\bibitem[{{Ferrand} {et al.}(2010)}]{Ferrand-etal-2010} {Ferrand G. et al. 2010, A\&A, 509, L10}
\bibitem[{{Fraschetti}(2013)}]{Fraschetti-2013} Fraschetti, F. 2013, ApJ 770, 84
\bibitem[{{Fulbright} \& {Reynolds}(1990)}]{Fulbright-Reynolds-1990} Fulbright M., Reynolds S. 1990, ApJ 357, 591
\bibitem[{{Giacalone} \& {Jokipii}(2007)}]{Giacalone-Jokipii-2007} Giacalone J., Jokipii J. 2007, ApJ 663, L41
\bibitem[{{Guo} {et al.}(2012)}]{Guo-et-al-2012} Guo F., Li S., Li H., Giacalone J., Jokipii J., Li D., 2012, ApJ 747, 98 
\bibitem[{{Hnatyk} \& {Petruk}(1996)}]{Hn-Pet-96} {Hnatyk} B., {Petruk} O. 1996, Kinematics Phys. Celest. Bodies, 12, 35
\bibitem[{{Hnatyk} \& {Petruk}(1999)}]{Hn-Pet-99} {Hnatyk} B., {Petruk} O. 1999, A\&A, 344, 295
%\bibitem[{{Korobeinikov}(1991)}]{Korob1991} Korobeinikov V., Problems of Point Blast Theory (Springer-Verlag New York, 1991), 400 p.
\bibitem[{{Kulsrud} \& {Cesarsky}(1971)}]{Kulsrud-Cesarsky-1971} Kulsrud R., Cesarsky C., 1971, Astrophys. Lett. 8, 189
\bibitem[{{Lagage} \& {Cesarsky}(1983)}]{Lagage-Cesarsky-1983} {Lagage P., Cesarsky C. 1983, A\&A 125, 249}
\bibitem[{{Laming}(2015)}]{Laming-2015} Laming J. M. 2015, ApJ 805, 102
\bibitem[{{Long} {et al.}(2003)}]{Long-et-al-2003} Long K. et al., 2003, ApJ 586, 1162
\bibitem[{{Malkov} {et al.}(2012)}]{Malkov-et-al-2012} Malkov M., Sagdeev R., Diamond P. 2012, ApJ 748, L32
\bibitem[{{McKenzie} \& {V\"olk}(1982)}]{McKenzie-Volk-1982} McKenzie J., V\"olk H., 1982, A\&A 116, 191
\bibitem[{{Miceli} {et al.}(2009)}]{Miceli-et-al-2009} Miceli M., Bocchino F., Iakubovskyi D., Orlando S., Telezhinsky I., Kirsch M., Petruk O., Dubner G., Castelletti G. 2009 A\&A 501, 239
\bibitem[{{Mizuno} {et al.}(2011)}]{Mizuno-et-al-2011} Mizuno Y., Pohl M., Niemiec J., Zhang B., Nishikawa K.-I., Hardee P. 2011 ApJ 726, 62
%Mizuno et al. 2014 MNRAS, 439, 3490
\bibitem[{{Morlino} {et al.}(2010)}]{Morlino-et-al-2010} Morlino G., Amato E., Blasi P., Caprioli D. 2010, MNRAS Let. 405, L21
%\bibitem[{{Morlino} \& {Caprioli}(2012)}]{Morlino-Caprioli-2012} Morlino G., Caprioli D., 2012, A\&A 538, A81
\bibitem[{{Orlando} {et al.}(2007)}]{Orlando-etal-2007} Orlando S., Bocchino F., Reale F., Peres G., Petruk O., 2007, A\&A 470, 927
\bibitem[{{Orlando} {et al.}(2011)}]{Orlando-etal-2011} Orlando S. Petruk O., Bocchino F., Miceli M., 2011, A\&A 526, A129
\bibitem[{{Orlando} {et al.}(2012)}]{Orlando-etal-2012} Orlando S. Bocchino F., Miceli M., Petruk O., Pumo M. 2012, ApJ 749, 156
\bibitem[{{Petruk}(2000)}]{approxSedov} Petruk O. 2000, A\&A 357, 686
\bibitem[{{Petruk}(2001)}]{petruk-2001} Petruk O. 2001, A\&A 371, 267
\bibitem[{{Petruk} {et al.}(2009a)}]{Petruk-et-al-2009} Petruk O., Dubner G., Castelletti G. et al. 2009a, MNRAS 393, 1034
\bibitem[{{Petruk} {et al.}(2009b)}]{Petruk-et-al-2009b} Petruk O., Beshley V., Bocchino F., Orlando S. 2009b MNRAS  395, 1467
\bibitem[{{Petruk} {et al.}(2009c)}]{Petruk-etal-2009c} Petruk O., Bocchino F., Miceli M., Dubner G., Castelletti G., Orlando S., Iakubovskyi D., Telezhinsky I. 2009c, MNRAS 399, 157
\bibitem[{{Petruk} {et al.}(2011a)}]{Petruk-et-al-2011} Petruk O. et al. 2011a, MNRAS 413, 1643
\bibitem[{{Petruk} {et al.}(2011b)}]{xmaps} Petruk O., Orlando S., Beshley V., Bocchino F., 2011b, MNRAS 413, 1657
\bibitem[{{Petruk} {et al.}(2012)}]{Petruk-etal-2012} Petruk O., Kuzyo T., Bocchino F. 2012 MNRAS 419, 608
\bibitem[{{Petruk} {et al.}(2016)}]{petr-2016} Petruk O., Kuzyo T., Beshley V., 2016, MNRAS, 2016, MNRAS 456, 2343
\bibitem[{{Pohl} {et al.}(2005)}]{Pohl-et-al-2005} Pohl M., Yan H., Lazarian A. 2005, ApJ 626, L101
\bibitem[{{Pohl} {et al.}(2015)}]{Pohl-et-al-2015} Pohl M. Wilhelm A., Telezhinsky I. 2015 A\&A 574, A43 
\bibitem[{{Ptuskin} \& {Zirakashvili}(2003)}]{Ptuskin-Zirak-2003} Ptuskin V., Zirakashvili V., 2003, A\&A 403, 1
\bibitem[{{Rakowski} {et al.}(2011)}]{Rakowski-et-al-2011} Rakowski C. et al 2011 ApJ 735 L21
\bibitem[{{Ressler} {et al.}(2014)}]{Ressler-et-al-2014} Ressler S., Katsuda S., Reynolds S., Long K., Petre R., Williams B., Winkler P. F. 2014 ApJ 790, 85
\bibitem[{{Reynolds}(1998)}]{Reyn-98} Reynolds S., 1998, ApJ 493, 375
\bibitem[{{Reynolds}(2004)}]{Reynolds-2004} Reynolds S. 2004 Adv. Sp. Res. 33, 461
\bibitem[{{Tran} {et al.}(2015)}]{Tran-et-al-2015} Tran A., Williams B., Petre R., Ressler S., Reynolds S. 2015, ApJ 812, 101
\bibitem[{{Sedov}(1959)}]{Sedov-59} Sedov L., 1959, {Similarity and Dimensional Methods in Mechanics} (New York, Academic Press).
\bibitem[{{Schneiter} {et al.}(2015)}]{Schneiter-etal-2015} Schneiter E., V\'elazquez P., Reynoso E., Esquivel A., de Colle F., 2015, MNRAS 449, 88
\bibitem[{{Skilling}(1975a)}]{Skilling1975i} Skilling J., 1975a, MNRAS 172, 557 
\bibitem[{{Skilling}(1975b)}]{Skilling1975iii} Skilling J., 1975b, MNRAS 173, 255 
\bibitem[{{Stroman} \& {Pohl}(2009)}]{Stroman-Pohl-2009} Stroman W., Pohl M. 2009 ApJ 696, 1864
\bibitem[{{Uchiyama} {et al.}(2007)}]{Uchiyama-et-al-2007} Uchiyama Y., Aharonian F., Tanaka T., Takahashi T., Maeda Y. 2007, Nature 449, 576
\bibitem[{{Vink} \& {Laming}(2003)}]{vink-laming-2003} {Vink J., Laming J. 2003, ApJ 584, 758} 
\bibitem[{{V\"olk} {et al.}(2005)}]{Volk-Ber-2005-mf} {V\"olk H., Berezhko E., Ksenofontov L. 2005, A\&A 433, 229}
\bibitem[{{Warren} {et al.}(2005)}]{Warren-et-al-2005} Warren J., Hughes J., Badenes C., Ghavamian P. et al. 2005, ApJ 634, 376
\bibitem[{{West} {et al.}(2016)}]{West-SafiHarb-et-al-2016} West J., Safi-Harb S., Jaffe T., Kothes R., Landecker T., Foster T. 2016, A\&A 587, A148
\bibitem[{{Winkler} {et al.}(2014)}]{Winkler-etal-2014} {Winkler P., Williams B., Reynolds S., Petre R., Long K., Katsuda S., Hwang U. 2014, ApJ, 781, 65}
\bibitem[{{Woltjer}(1970)}]{Woltjer-1970} {Woltjer L. 1970 Proc. IAU Symposium 39, 229} 
\end{thebibliography}
\end{document}